\documentclass[fleqn,usenatbib]{rasti}

\usepackage{newtxtext,newtxmath}

\usepackage[T1]{fontenc}
\usepackage{multicol}
\usepackage{multirow}
\usepackage{widetext}
\usepackage{graphicx}
\usepackage{float}
\usepackage{lipsum}

\DeclareRobustCommand{\VAN}[3]{#2}
\let\VANthebibliography\thebibliography
\def\thebibliography{\DeclareRobustCommand{\VAN}[3]{##3}\VANthebibliography}

\usepackage{graphicx}	
\usepackage{amsmath}	

\title[E(2)-Equivariant Features for Radio Galaxy Classification]{E(2)-Equivariant Features in Machine Learning for Morphological Classification of Radio Galaxies}

\author[N.~E.~P.~Lines et al.]{
Natalie E.~P.~Lines,$^{1}$\thanks{Equal contribution: natalie.lines@port.ac.uk}
Joan Font-Quer Roset,$^{1}$\thanks{Equal contribution: joan.fontquer.roset@gmail.com}
and Anna M.~M.~Scaife$^{1,2}$
\\
$^{1}$Jodrell Bank Centre for Astrophysics, Department of Physics \& Astronomy, University of Manchester, Oxford Road, Manchester M13 9PL UK\\
$^{2}$The Alan Turing Institute, Euston Road, London, NW1 2DB, UK\\
}

\date{Accepted XXX. Received YYY; in original form ZZZ}

\pubyear{2022}

\begin{document}
\label{firstpage}
\pagerange{\pageref{firstpage}--\pageref{lastpage}}
\maketitle

\begin{abstract}
With the growth of data from new radio telescope facilities, machine-learning approaches to the morphological classification of radio galaxies are increasingly being utilised. However, while widely employed deep-learning models using convolutional neural networks (CNNs) are equivariant to translations within images, neither CNNs nor most other machine-learning approaches are equivariant to additional isometries of the Euclidean plane, such as rotations and reflections. Recent work has attempted to address this by using $G$-steerable CNNs, designed to be equivariant to a specified subset of 2-dimensional Euclidean, E(2), transformations. Although this approach improved model performance, the computational costs were a recognised drawback. Here we consider the use of directly extracted E(2)-equivariant features for the classification of radio galaxies. Specifically, we investigate the use of Minkowski functionals (MFs), Haralick features (HFs) and elliptical Fourier descriptors (EFDs). We show that, while these features do not perform equivalently well to CNNs in terms of accuracy, they are able to inform the classification of radio galaxies, requiring $\sim$50 times less computational runtime. We demonstrate that MFs are the most informative, EFDs the least informative, and show that combinations of all three result in only incrementally improved performance, which we suggest is due to information overlap between feature sets.
\end{abstract}



\section{Introduction}

Radio galaxies are galaxies with strong radio emission, typically forming jets and lobes on either side of the host galaxy, which originate from the synchrotron emission of active galactic nuclei (AGNs). The morphologies of these radio galaxies were first classified into two types by \cite{FRpaper}, known as FR type I (FR\,I) and type II (FR\,II). 
These two classes are based on the relative locations of highest surface brightness; in FR\,I, the brightest area lies towards the centre of the AGN, known as edge-darkened, while in FR\,II the brightest regions lie further away from the centre of the galaxy, known as edge-brightened. These are classified based on their Fanaroff-Riley ratio $\rm R_{FR}$, defined as the ratio of the distance between the two brightest regions located on opposing sides of the galaxy to the total extension of the galaxy at the lowest threshold. FR\,I galaxies are those with $\rm R_{FR}<0.5$, and FR\,II are those with $\rm R_{FR}>0.5$. In the original investigation by \cite{FRpaper}, these two types were shown to be divided in power: FR\,I radio galaxies typically have powers at 1.4 GHz below $P_{1.4GHz}=10^{25}\,{\rm W\,Hz^{-1}}$, while FR\,II galaxies typically have powers above this. Understanding the formation of these different radio galaxies, the accretion process and interactions of the jets and surrounding environment in the galaxies remains an active area of research, and requires large numbers of FR type labelled radio galaxies \citep{hardcastle2020}.

Upcoming telescopes are expected to detect radio galaxies in the millions, with the Square Kilometre Array (SKA) expected to detect upwards of 500 million radio sources \citep{500mil}, and the Evolutionary Map of the Universe project (EMU) expected to find 70 million radio sources \citep{70mil}, of which around 10\% are estimated to be complex objects that will require cross-identification. Though attempts have been made to manually sort large datasets of galaxies, such as the Radio Galaxy Zoo project \citep{radio_galaxy_zoo}, automated radio galaxy classification is becoming more essential to allow for a full exploitation of the data available. 

\subsection{Machine Learning for Radio Galaxy Classification} 
\label{previous machine learning implementations}

Many different machine learning models have been implemented in astronomy for the task of radio galaxy classification. In a review of the recent works on radio galaxy morphological classification by \cite{review}, convolutional neural networks (CNNs) are shown to be the most popular machine learning method used. CNNs use convolutional layers to extract features from an image, which are integrated into a neural network. CNNs were first developed by \cite{fukushima_1980}, and are the standard tool for computational vision due to their ability to capture information stored in complex images. First examples of using neural networks for classification in astrophysics date back decades to works such as \cite{1992_ml} and \cite{1995_ml}, and they were first implemented into the task of FR classification in \cite{first_fr_cnn}. Since then, works such as that by \cite{lukic} and \cite{Ma_2019}, have built on this work, developing models with accuracies of 93--94\%.

In addition to CNNs, other machine learning models have been used for the problem of FR classifications. A review of different conventional machine learning methods has shown that random forests were the best of the tested models\footnote{These were: Nearest Neighbours, Linear Support Vector Machine (SVM), Radial Basis Function SVM, Gaussian Process Regression, Random Forest, Multi-layered Perceptron Neural Network, AdaBoosted Decision Tree, Naive Bayes and Quadratic Discriminant Analysis.} \citep{rfc_fr}, achieving an accuracy of 95\% using features such as lobe size and brightness as inputs to the model. Gradient boosting models such as XGBoost have been found to perform similarly to random forests \citep{gradient_boosted_fr}.

However, many of these models have drawbacks. CNNs extract information from images through the use of kernels, which contain learnable weights, allowing CNNs to be adaptable to learning different features. This requires the training of more parameters which is less computationally efficient, and although recent work has been done to reduce the number of parameters needed while maintaining similar performance levels \citep[see e.g.][]{micah_0.5_less_params}, \textcolor{black}{there is also wider interest in developing more computationally efficient approaches, such as the trainable COSFIRE descriptor \citep{cosfire}}. Another drawback to CNNs is that although they are translationally equivariant, the convolutional steps mean that they are not intrinsically equivariant to rotations and reflections, and as such two images of the same galaxy at different orientations are not necessarily classified consistently. When training models on augmented datasets, in which the dataset includes multiple transformed copies of the same image, the model is required to independently classify the same object at different orientations. Data augmentation has been used in works such as that by \cite{data_aug_eg}, and has been shown to increase performance under E(2) symmetries  \citep{data_aug_good}.

One data space solution to the non-equivariance of CNNs, used in the works by \cite{pink_polsterer} and \cite{rotation_standardised_fr}, involves normalising the orientation of the galaxies, which is shown to improve both the accuracy and the training time of the model \citep{rotation_standardised_fr}. A model space solution is the use of $G$-steerable CNNs \citep{cnn_equivariance}, in which the desired symmetries are encoded into the weight sharing of the kernel, ensuring mathematical equivariance under these symmetries. This has been shown to be more effective than data augmentation, and has been implemented into the problem of FR classification by \cite{E2CNN}, where the accuracy is found to improve from 94\% for a conventional CNN to 97\% for a $G$-steerable CNN, though at the cost of increased computational time. 

In this work we test alternative methods of feature extraction that are equivariant under rotations, reflections and translations to replace the convolutional steps of a standard CNN when classifying radio galaxies according to their FR type. The structure of this paper is as follows: in Section~\ref{sec:equitheory} we define the concepts of isometric invariance and equivariance, including how these can be built in to convolutions; in Section~\ref{sec:features} we describe the equivariant feature sets used in this work; in Section~\ref{sec:data} we introduce the data set that will be used, including the pre-processing steps that are applied, and in Section~\ref{sec:ml} we outline the machine learning methods that are employed, including model training and hyperparameter tuning details. In Section~\ref{sec:performance}, we report on the performance of the trained models for different feature set combinations, and in Section~\ref{sec:discussion} we evaluate these results and discuss the importance of different features, with additional investigation of the representation space for each feature set using dimensionality reduction. In Section~\ref{sec:conclusion} we draw our conclusions.

\section{Isometric Equivariance and Invariance} 
\label{sec:equitheory}

Here we introduce a simplified explanation of the concepts of equivariance and invariance  for Euclidean transformations including the dihedral and cyclic subgroups of the orthogonal group $\mathrm{O}(n)$, following \cite{Weyl:1980}. For the $n$-dimensional real space $\mathbb{R}^{n}$, equivariance is described as follows: let $\phi$ be a mapping $\phi \colon X \longrightarrow Y$, where $X$ and $Y$ are subsets of the real space, $X, Y\subseteq\mathbb{R}^{n}$. Suppose $G$ is a group of transformations acting on $X$ and $Y$. If $\phi$ commutes with each element of $G$ for all elements of $X$, i.e.
\begin{equation} \label{equivariance}
\color{black}{[\phi \cdot \rho (g)](x) = [\rho(g) \cdot \phi](x)} \quad \forall~g\in G,~x\in X,
\end{equation}
then $\phi$ is said to be equivariant with respect to $G$\textcolor{black}{, where $\rho(g)$ is the representation of the action $g$ in $\mathbb{R}^{n}$}. If instead
\begin{equation} \label{invariance} 
\color{black}{[\phi \cdot \rho(g)](x)} = \phi(x) \quad \forall~g\in G,~x\in X,
\end{equation}
then $\phi$ is said to be invariant with respect to $G$. 

The $n$-dimensional Euclidean group, E$(n)$, is the set of transformations that preserve Euclidean distance, and comprises of translations, rotations and reflections. Given $r\in \mathbb{R}^{n}$, a translation by $r$ is defined by the map
\begin{equation} \label{translation}
    t_{r}:\mathbb{R}^{n} \longrightarrow \mathbb{R}^{n},\quad t_{r}(x)=x+r.
\end{equation}
The set of $n$-dimensional vectors with real coefficients under vector addition, $(\mathbb{R}^{n}, +)$, is the group of all translations.

The group of non-translational distance-preserving transformations of the $n$-dimensional Euclidean space is the orthogonal group, O$(n)$, defined as $\mathrm{O}(n)=\{ A\in \mathrm{GL}(n) \mid A^{-1}A=AA^{-1}=\mathbb{I} \}$, where $\mathrm{GL}(n)$ is the set of invertible $n\times n$ matrices. In order to preserve distances, the magnitude of the determinant of these matrices has to equal $1$, and as such the orthogonal group is split into two components, those of matrices with determinant $1$, and those of matrices with determinant $-1$.

The set of matrices in O($n$) with a determinant of $1$ forms a subgroup, known as the special orthogonal group, $\mathrm{SO}(n)=\{ A \in \mathrm{O}(n) \mid \mathrm{det}(A)=1 \}$. In two dimensions, the special orthogonal group is the group of all rotations of the plane. For example, a rotation by an angle $\alpha$ of a set of points $(x_n,\ y_n)\in\mathbb{R}^2$ is expressed as 
    \begin{equation} \label{rotationmatrix}
        \begin{bmatrix}x_{n}'\\ y_{n}'
        \end{bmatrix}
        =
        \begin{bmatrix}\cos({\alpha}) & - \sin({\alpha})\\ \sin({\alpha}) & \cos({\alpha})
        \end{bmatrix}\begin{bmatrix}
        x_{n}\\ y_{n}
        \end{bmatrix}.
    \end{equation}

The set of orthogonal matrices of determinant $-1$, which is given by $\mathrm{O}(n) \setminus \mathrm{SO}(n) = \{ A \in \mathrm{O}(n) \mid \mathrm{det}(A)=-1 \}$, represents reflections in two dimensions. A reflection by an angle $\beta$ of a set of points $(x_n,\ y_n)\in\mathbb{R}^2$ is written as
    \begin{equation} \label{reflectionmatrix}
        \begin{bmatrix}x_{n}'\\ y_{n}'
        \end{bmatrix}
        =
        \begin{bmatrix}\cos({2\beta}) & \sin({2\beta})\\ \sin({2\beta}) & - \cos({2\beta})
        \end{bmatrix}\begin{bmatrix}
        x_{n}\\ y_{n}
        \end{bmatrix}.
    \end{equation}
Consequently, every element of $\mathrm{O}(n)$ is either a rotation or a reflection, and the Euclidean group E$(n)$ can be expressed as the semi-direct product E$(n) = (\mathbb{R}^{n}, +) \rtimes \mathrm{O}(n)$. Therefore, for a map to be equivariant to all Euclidean isometries, it must be equivariant with respect to $(\mathbb{R}^{n}, +)$ and $\mathrm{O}(n)$. Commonly used subgroups of E$(2)$ are the cyclic group $C_{n}$, which contains all the rotations by multiples of $\frac{2\pi}{n}$, and the dihedral group $D_{n}$, which contains all the rotations of $C_n$ in addition to $n$ reflections.

\subsection{Group-Equivariant Convolutions} \label{g-equivariant convolutions} 

The convolutional layers of a CNN convolve the input image with a kernel of weights, which are parameters learned by the CNN. Convolutions involve taking the dot product between the kernel and every subsection of the image of the same shape as the kernel, producing a feature map. As this process involves applying the kernel systematically to every subsection of the image, the weights of the kernel are shared, meaning this process is equivariant to translations in the input image up to edge effects.  This translational equivariance is expressed as
\begin{equation}
\left[\left[\rho\left(g_{\rm trans}\right) \cdot f\right] \star \psi\right](x)=\left[\rho\left(g_{\rm trans}\right) \cdot\left[f \star \psi\right]\right](x)
\end{equation}
for a kernel $\psi$, translation $\rho(g_{\rm trans})$ and image $f:\mathbb{R}^2 \rightarrow \mathbb{R}$, where $\star$ represents convolution. Convolutions are linear operations, so an activation function is applied to the feature map to introduce non-linearity. Following this, a pooling layer is typically used to decrease the size of the feature maps. This is done by dividing the feature map  into equally sized subsections and representing each subsection by a single value such as the mean or the maximum of the pixels in the subsection. This allows the representation to be invariant to small translations in the input, and also increases computational efficiency due to the down sampling. Though equivariant to translations, convolutions are not equivariant to rotations and reflections. Instead, the convolution of a rotated image, $\rho\left(g_{\rm rot}\right) \cdot f$, with a kernel is equivalent to convolving the original image with the inverse-rotated kernel, $\rho\left(g_{\rm rot}^{-1}\right) \cdot \psi$, and then rotating the output, expressed by
\begin{equation} 
\label{eq:rotational inequivaraince}
    \left[ \left[\rho\left(g_{\rm rot}\right) \cdot f\right] \star \psi \right](x)=\left[ \rho\left(g_{\rm rot}\right) \cdot\left[f \star \left[ \rho\left(g_{\rm rot}^{-1}\right) \cdot \psi\right]\right]\right](x)
\end{equation}
\citep{cnn_equivariance}.

Equivariance of convolutions under other symmetry transformations can be imposed by restricting the form of the convolution kernel. Inspection of Equation \ref{eq:rotational inequivaraince} shows that this satisfies the definition of equivariance when the kernel, $\psi$, also encompasses $\rho(g^{-1}) \cdot \psi$.
\textcolor{black}{ 
Equivalently, we can speak of equivariance not just as convolving with one kernel but with multiple, i.e. if instead of convolving an image with a single kernel, we convolve it with a stack kernels that correspond to every possible rotation of the original kernel to produce a stack of feature maps, then this stack convolution will be equivariant with respect to rotation \citep{cnn_equivariance}.}
 This motivates the extension to the definition of the convolution operation for a group action $g \in G$ to be 
\begin{equation} \label{eq:convolution definition}
[f \star_g \psi](g)=\sum_{h \in H} \sum_k f_k(y) \psi_k\left(g^{-1}\cdot h\right),
\end{equation}
where $k$ is the input channel, $H=\mathbb{R}^2$ for the first layer (lifting convolutions) and $H=G$ for the following layers ($G$-convolutions). The result of this is a feature map of higher dimensionality than the input image due to the dependence of the definition of the convolution operation, $\star_g$, on the group action, $g$. When $G$ is the set of all translations, this simplifies to the standard definition of the convolution operation due to the commutativity of $g$ and $h$. Furthermore, to allow for equivariance, the kernel should satisfy
\begin{equation}
    \psi(gx)=\rho(g)\psi(x)\rho(g^{-1}),
\end{equation}
for all $g\in G$ and $x \in \mathbb{R}^2$.

\section{Equivariant Feature Extraction}
\label{sec:features}

\subsection{Minkowski Functionals}
\label{minkowski functionals}

\begin{figure}
    \centering
    \includegraphics[width=0.5\textwidth]{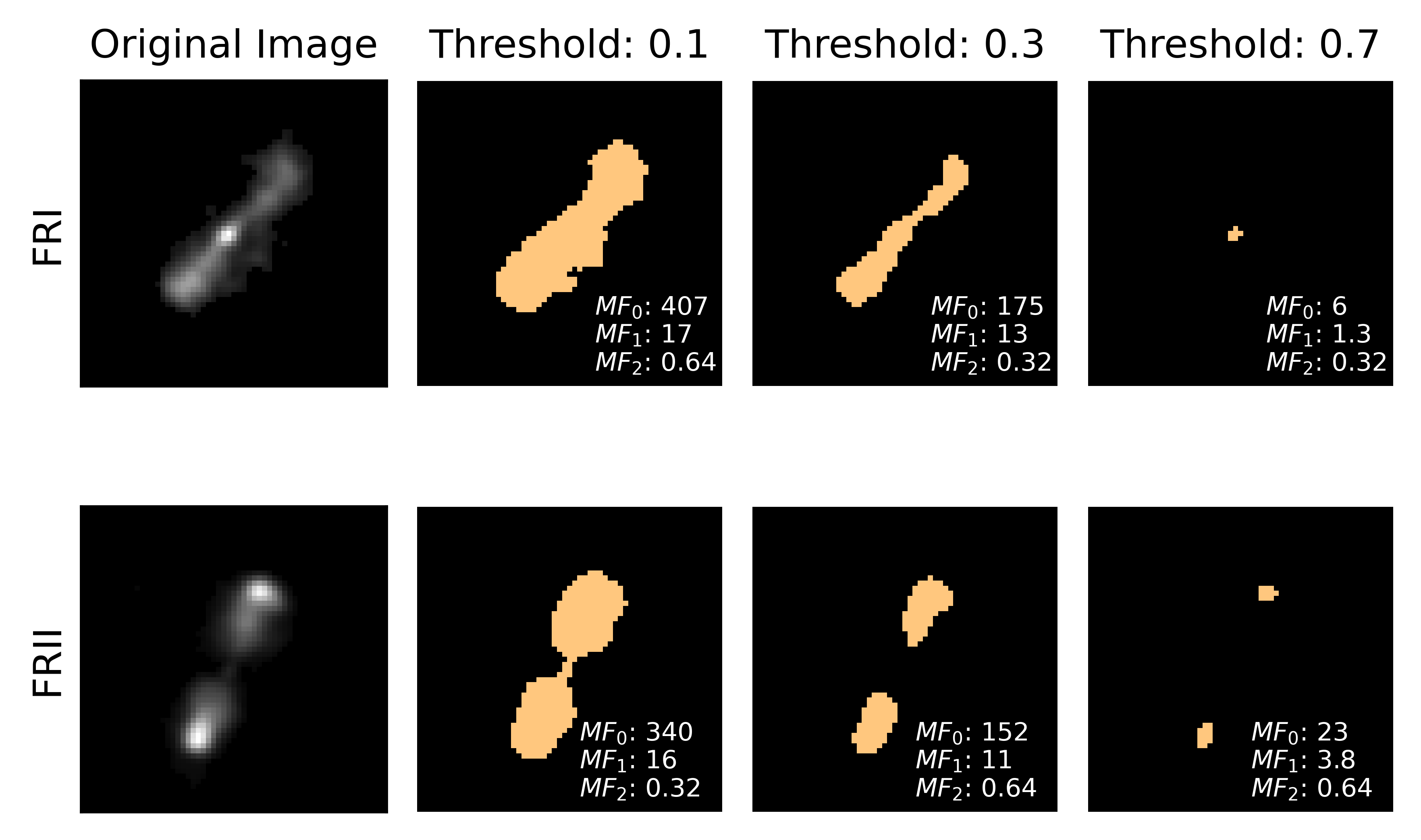}
    \caption{MFs of the binary images created by imposing a relative threshold of 0.1, 0.3 and 0.7 for a FR\,I and FR\,II galaxy. Values of the MFs are shown in the subplots, with the original galaxy images displayed on the left. As the threshold increases, MF 2 tends to the value of $\frac{1}{\pi}$ for the FR\,I galaxy and $\frac{2}{\pi}$ for the FR\,II galaxy.}
    \label{fig:mfs}
\end{figure}

Minkowski functionals (MFs) are a class of morphological descriptors of continuous fields, developed by \cite{Minkowksi}. They provide a generalisation of the smooth curvature integrals for the number of dimensions of the data, quantifying its shape and connectivity. Their previous applications primarily consist of statistical descriptors of non-Gaussian fields, such as spatial patterns in the large-scale structure of the universe \citep{mf_eg2}, weak lensing convergence maps \citep{mf_eg1}, galaxy clustering \citep{mf_og}, and the cosmic microwave background \citep{mf_eg3}, and are often used in the context of providing an alternative to the $n$-point correlation function. They are useful in these contexts due to their ability to capture information on larger scales than typical of 2-point correlation functions, and can encompass correlation at arbitrary orders.

Though previous works typically use MFs as topological descriptors of random fields, their applicability extends to any set whose boundary is smooth. MFs are invariant under E(2) symmetries, and for $n$ dimensions there are $n+1$ MFs. Furthermore, \cite{hadwiger} showed that a functional in $n$ dimensions is continuous and both translationally and rotationally invariant if and only if it is a linear combination of the $n+1$ MFs in that number of dimensions, making it a natural tool for our work. These $n+1$ MFs depend on a threshold, $\nu$, used to define the excursion set of points above the thresholds, converting the input to binary form.
Consequently, in two dimensions there are three MFs, proportional to area, perimeter, and Euler characteristic respectively. A full mathematical definition can be found in Appendix~\ref{app:mf}. The values of the MFs evaluated at three different thresholds for an FR\,I and FR\,II galaxy are shown in Figure~\ref{fig:mfs}.

\subsection{Haralick Features}
\label{haralick features}

\begin{figure}
    \centering
    \includegraphics[width=0.5\textwidth]{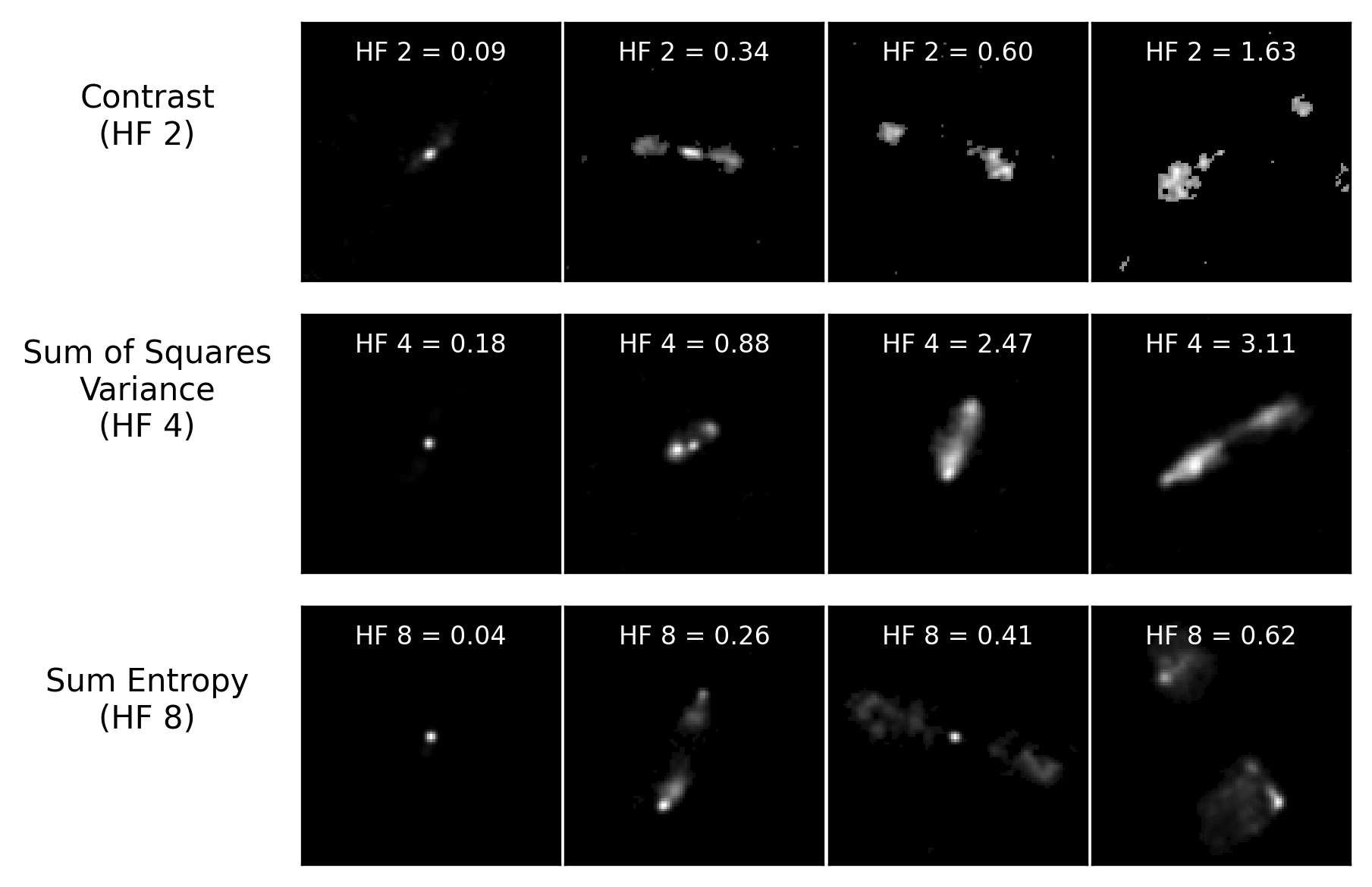}
    \caption{Example galaxies corresponding to increasing values of a specific HF. Top row shows galaxies of increasing contrast, the second HF. Middle row shows galaxies of increasing sum of squares variance, the fourth HF. Bottom row shows galaxies of increasing sum entropy, the eighth HF.}
    \label{fig:haralicks}
\end{figure}

Haralick features (HFs) are a set of features proposed by \cite{haralick_og} as mathematical descriptors of image texture, with the applicability to classification problems in mind. They are calculated from the Grey Level Co-occurrence Matrix (GLCM), which encodes information about the spatial distribution of variations in pixel brightness of an image. HFs have previously been implemented in astronomy by \cite{haralick_radio}, using these features to sort radio galaxies in an unsupervised setting, though their previous applications in the context of machine learning have primarily been in medical imaging for tasks such as cellular dose distributions \citep{haralick_medicine}, and tumour classification \citep{hf_med2}. There are 14 HFs that can be calculated from the GLCM, however these previous studies have found the 14$^{\rm th}$ feature to be numerically unstable so only the first 13 are typically used.

The GLCM contains textural information through encoding the relative pixel brightness of adjacent pixels. Unlike in convolutions, the resulting matrix does not preserve any topology of the image, but rather describes the frequency at which two pixels of certain brightness appear relative to each other. For an image where each pixel $(x_i,y_i)$ has a value $I(x_i,y_i)\in \{0, 1,\ldots, I_{\rm max}\}$, the resulting GLCM matrix, $P(a,b)$, will have shape $I_{\rm max} \times I_{\rm max}$. Two pixels are said to be related,  $(x_1,y_1) \sim (x_2,y_2)$, if they have a (user) specified relative positioning. For example: horizontally adjacent, in which case $(x_1, y_1)\sim (x_2,y_2) \iff$ $(y_1 = y_2$ and $x_1=x_2 \pm 1)$. 

The matrix $P(a,b)$ is then defined as 
\begin{equation}
    P(a,b)= \#\{(x_1,y_1)\sim (x_2, y_2)\:|\: I(x_1,y_1)=a\:,\:I(x_2,y_2)=b\},
\end{equation}
where for each matrix entry, its index is the pixel brightness being considered, and its value is the number of times two pixels of that brightness appear with the specified relative positioning. 

The process of calculating the matrix elements, $P(a,b)$, is invariant under translations and equivariant under reflections and rotations (e.g. the number of pixels horizontally adjacent in the image is the same as the the number of pixels vertically adjacent in the image after rotation by 90$^{\circ}$). Rotational and reflectional invariance of the features extracted from this matrix can be imposed by restricting the form the relation can take: reflectional invariance can be imposed by requiring the relation to be symmetric, $(i_1\sim i_2 \implies i_2 \sim i_1)$, which is an a priori assumption for any metric-like relation; rotational invariance can be imposed by requiring the relation to be independent of angle, or by calculating the HFs for all possible angles at a set distance and computing the average. From the GLCM, the 14 different textural features can be extracted, with each of these quantifying a different textural property such as contrast, variance, and entropy, as shown in Figure~\ref{fig:haralicks}. A list of all HFs along with their mathematical definition is given in Appendix~\ref{app:haralicktheory}.

\subsection{Elliptical Fourier Descriptors}\label{elliptical fourier descriptors}

Elliptical Fourier analysis was formalised by \cite{efd_og}, and provides a tool to describe complex contours through approximating a contour as sums of sine and cosine functions. Previous uses include morphometry \citep{morphom1}, describing lake morphologies on Titan \citep{titan},  and the Hubble classification of galaxies \citep{efd_galaxy}. In normal Fourier analysis, a contour in two dimensions is converted into a single curve in the frequency domain which is then approximated, treating the two dimensions simultaneously. This simultaneous parametrisation means this analysis is unable to fit to any contour which folds back on itself with respect to the polar angle, limiting the applicability to polar convex shapes. In elliptical Fourier analysis, the $x$ and $y$ coordinates are reconstructed in the frequency domain independently, where the harmonics are ellipses. It is these Fourier coefficients, $\{A_n, B_n, C_n, D_n\}$, that are referred to as the elliptical Fourier descriptors (EFDs) of order $n$. 

In two dimensions, the $x$-coordinates and $y$-coordinates are separately decomposed as
\begin{equation}
    y_N=A_0+\sum_{n=1}^N A_n \cos (n t)+\sum_{n=1}^N B_n \sin (n t), ~~{\rm and}
\end{equation}
\begin{equation}
    x_N=C_0+\sum_{n=1}^N C_n \cos (n t)+\sum_{n=1}^N D_n \sin (n t),
\end{equation}
where the maximum order $N$ sets a limit on how close the approximation is to the true contour, and $t \in \left[0, 2\pi \right)$. More mathematical detail on these features can be found in Appendix~\ref{app:efdtheory}.

\begin{figure}
    \centering
    \includegraphics[width=0.5\textwidth]{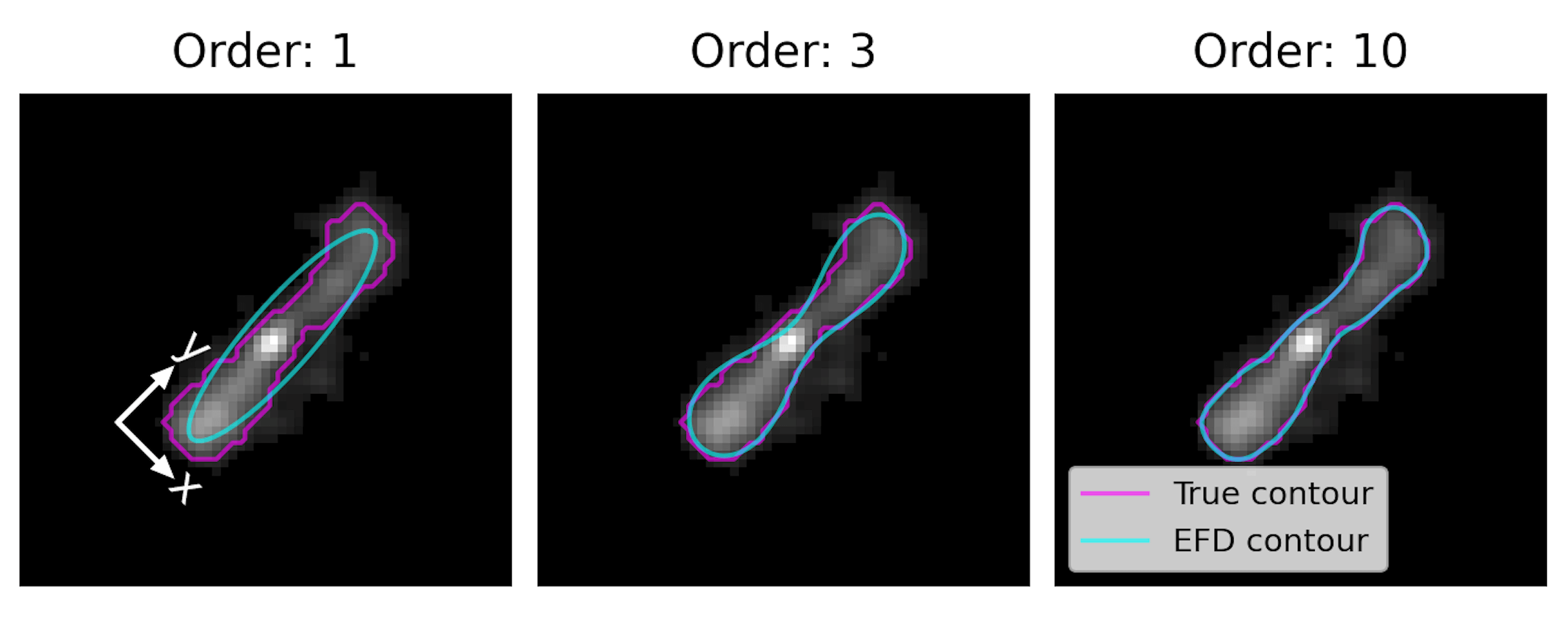}
    \caption{Elliptical Fourier analysis of the contour of a FR\,I galaxy at threshold = 0.2. Shown in purple is the true contour at the threshold, and shown in blue are the contours using the EFDs at order 1, 3 and 10. To allow for direct comparison of values of the individual EFDs, these are calculated with respect to the coordinates shown in the left-most image.}
    \label{fig:efd}
\end{figure}

Due to the coordinate-dependence of these functions, EFDs are not invariant under E(2) symmetries. They are rotationally and reflectionally equivariant, and the translational information is captured in the constants $A_0$ and $C_0$, about which the contour is centred. Many implementations of EFDs impose invariance to allow for direct comparison of the EFDs by rotating the image to align the major axis of the ellipse at $n=1$ to the $y$-axis, as shown in Figure~\ref{fig:efd}. They can further be made scale invariant by normalising the length of the major axis. As with MFs, calculation of EFDs requires a threshold, $\nu$, to be defined, such that the curve being approximated is the boundary of the excursion set, $\Sigma(\nu)$.  

\section{Data} 
\label{sec:data}

\subsection{MiraBest} 

MiraBest is a catalogue of labelled radio galaxies compiled by \cite{mirabest}. It comprises of data from data release 7 \citep{dr7}  of the Sloan Digital Sky Survey (SDSS) \citep{sdss}, the Northern Very Large Array (VLA) Sky Survey (NVSS) \citep{nvss}, and the Faint Images of the Radio Sky at Twenty centimetres (FIRST) survey \citep{first}, cross-matched by \cite{best&heckman}. With the aim of these galaxies being morphologically classifiable, the dataset is restricted to extended galaxies containing AGN, and a 40 mJy flux cut-off is imposed to exclude extended galaxies that are too noisy to classify. \cite{mirabest} manually classified all of these galaxies into types FR\,I, FR\,II, hybrid, and unclassifiable. The hybrid class contains radio galaxies that would be classified as FR\,I according to one side of the galaxy but would be classified as FR\,II according to the opposite side, while the unclassifiable galaxies showed no strong characteristics of either FR\,I or FR\,II. In addition to the morphological type, each galaxy was given a label of either `confident' or `uncertain' based on how sure they were of the classification, and a sub-morphology out of standard, double-double, wide-angle tail, diffuse, or head-tail. The final catalogue contains 1329 sources.

In this work we use a version of the MiraBest dataset processed by \cite{mirabest_2} for the context of machine learning. As part of this processing, the unclassifiable radio galaxies were removed from the dataset, as well as any galaxies larger than a 150 $\times$ 150 pixel grid, any images that contained no data, and one galaxy that was in a morphological subclass on its own. The images had background noise removed, were centred, cropped to a 150 $\times$ 150 pixel grid (corresponding to 270 arcsecs for images from the FIRST survey), and the pixel values were normalised from 0 to 255. Additionally, the data was organised into 8 batches\footnote{These batches are built into the dataset, and are not the same as the batches used in the neural network training as described in Section~\ref{sec:nn_method} which we tune as a hyperparameter.}, with one labelled as a reserved test batch. The final dataset contains 1256 radio galaxies, of which 833 are labelled as confidently classified as either FR\,I (containing 397 samples) or FR\,II (containing 436 samples) with any sub-morphology. 

\subsection{Data Pre-processing} 
\label{data pre-processing}

In this study, we use the subset of 833 confidently classified FR\,I and FR\,II galaxies, known as `MBFRConfident', which is split into training $(\textcolor{black}{87.5\%})$ and testing $(12.5\%)$ data. Including only FR\,I and FR\,II and ignoring sub-morphologies simplifies the classification problem to a binary one, and as the FR\,I and FR\,II types make up the majority of the data, this avoids complexities with adding extra classes with an unbalanced split. Using only confidently labelled galaxies further simplifies the problem as it means that during the training process, performance on each galaxy can be treated equally. By including uncertain data, either the machine learning model would learn from its ability to classify confident and uncertain galaxies equivalently, which is undesirable especially in the case where the the uncertainty of the label potentially corresponds to a misclassification, or we would have to introduce a mechanism through which its predictions against confident and uncertain galaxies are treated differently. Though the full dataset of both confident and uncertain galaxies provides a better representation of radio galaxies as a whole, we also want to ensure that our model is able to deal with the simplified case before extending its adaptability. 

In machine learning, an increased convergence rate can be obtained through the process of normalising the input data such that the values are close to zero with a standard deviation of 1  \citep{backprop}. Following this, we normalise the images using $z$-score normalisation by 
\begin{equation}
\rm{normalised\; pixel = \frac{pixel - mean}{standard\; deviation}}.
\end{equation}
For the MBFRConfident dataset, the mean is 0.0031 and the standard deviation is 0.0350 \citep{mirabest_2}.  The only other pre-processing applied is for the calculation of HFs, as this requires an image of integer pixel values which we obtain through rounding. As the range of this background-removed data is -0.09 to 28.48, this results in a GLCM matrix of size 29 $\times$ 29, which is able to store adequate information to calculate the HFs at a relatively high precision while not impeding computational efficiency. 

\section{Machine learning implementation}
\label{sec:ml}

\subsection{Neural Network} 
\label{sec:nn_method}

To allow for a comparison, our neural network architecture follows on from \cite{E2CNN}, which is based on a LeNet \citep{Lenet_1998} structure, consisting of two convolutional and pooling layers followed by three fully connected layers. In our model, we replace the two convolutional layers with the feature calculation process, which contains no learnable parameters. Each of the three fully connected layers are followed by an activation function, which is ReLU in the case of the two hidden layers and softmax for the output layer. The input size of the first layer is the number of features being used, the final fully connected layer has size 2, and the two hidden layers have an unspecified number of neurons, which is varied during the hyperparameter tuning. Dropout is implemented before the final layer, with the dropout probability also being a hyperparameter. We use the cross-entropy loss function, and an Adam \citep{adam} optimiser to modify the learning rates, with the values of initial learning rate and weight decay as hyperparameters. This optimiser has been shown to be one of the best performing gradient descent optimisation algorithms \citep{optimiser_review}. The test data is  pre-defined, making up 12.5\% of the MBFRConfident dataset, and we split the remaining data such that the training data is 70\% of the full dataset and the validation data is the remaining 17.5\%, as this has been found to produce the best performance \citep{7030}. We run each of the features individually as well as together to gain an understanding of the information content of the features and to allow for a comparison. Each run has 1000 epochs, with early stopping implemented to save the best model which is used in testing, and we implement shuffling of the input data each epoch to avoid the model memorising the data. 

\textcolor{black}{Following hyper-parameter selection, as described further in Section~\ref{tuning of supervised models}, we used 10 realisations of the model, each trained using a randomised train/validation split and randomised initial parameter values, evaluated on the reserved test set. From these runs,} the mean and standard deviation of the performance of the model are calculated. 

Seven unique models types were trained: three for each individual feature type, three for each possible combination of features, and a final model that combines all three feature set extraction methods. All neural network models were built using the \texttt{PyTorch} \citep{pytorch} machine learning framework.

\subsection{XGBoost} 
\label{xgboost method}

\textcolor{black}{Decision trees form the basis of many machine learning algorithms; however, while 
individual decision trees are conceptually simple and can be applied to non-linear problems,  they are very unstable to changes in  training data and are strongly prone to overfitting \citep[see e.g.][]{Louppe}. To mitigate these effects, these decision trees can be combined through ensemble learning to produce more powerful models.}

\textcolor{black}{One such ensemble learning algorithm is a random forest. This involves creating an ensemble of decision trees, each of which is constructed on a bootstrapped sample of the datapoints in the dataset, referred to as bagging. Random forests extend this by adding a level of bootstrap sampling to the features, from which the decision nodes are constructed. This allows for better generalisation as it prevents overfitting and forces the model to not rely heavily on any one feature. Random forests often outperform neural networks for classification tasks on lower dimensional, less complex data. They have the benefit over neural networks that they tend to overfit less, are less computationally expensive, and perform reasonably well without the requirement of large training datasets. Their simplicity compared to neural networks also allows for better interpretability, which can provide insight into the features of the data.} 

\textcolor{black}{The alternative to bagging is gradient boosting, developed by \cite{gradient_boosting_og}, in which decision trees are constructed sequentially through learning the mistakes of the previous decision trees. The first instances of gradient boosting used an algorithm known as adaptive boosting \citep[Adaboost;][]{adaboost}, from which the more flexible gradient boosting algorithm was developed.}

Extreme Gradient Boosting \citep[XGBoost;][]{Chen:2016} is a tree ensemble machine learning model that can be trained to predict class labels. \textcolor{black}{XGBoost is a type of gradient boosting that is optimised for speed, computational efficiency,  and performance through integration of features such as parallelisation. The XGBoost algorithm follows the same structure as the one previously described, with an additional regularisation term added to the loss, similar to the weight decay parameter for neural networks. XGBoost tends to outperform random forests, but are prone to overfitting if hyperparameters are not tuned properly. Some of these key hyperparameters are max depth of decision trees, number of iterations, learning rate, and loss function.} 

\textcolor{black}{The training set-up for XGBoost is similar to that of the neural network} but is trained across a varying number of \textcolor{black}{decision} trees instead of a set number of epochs, and is tested on all seven of the possible combinations of features. We again use cross-entropy loss, with no added class weights due to the balance of the dataset. The training, validation and test splits are the same as that for the neural network, and for testing we again train the model with a different initial setup 10 times. XGBoost provides feature importances, which we average from testing the model 10 times. 

\subsection{Hyperparameter Tuning}
\label{tuning of supervised models}

The performance of neural networks and XGBoost depends on the values of their hyperparameters, and hence optimisation of these is essential. The neural network hyperparameters are number of thresholds, order, layer size, dropout probability, initial learning rate and weight decay. The XGBoost hyperparameters are number of thresholds, order, number of estimators and learning rate. Details of priors and optimal values are given in Appendix~\ref{appendix: hyperparameter tuning details}.

The space of all possible hyperparameters is very large and as the optimal values of hyperparameters are not independent, careful consideration is needed to be able to ensure an optimal set of hyperparameters is found. For this, we implement Bayesian optimisation using the developer platform Weights \& Biases \citep{wandb}, with at least 100 runs being executed before a set a hyperparameters were chosen. Bayesian optimisation searches the space of all possible hyperparameters in a way that is informed by the performance of previous runs. It constructs a probability distribution of the objective function in the hyperparameter space, and then tests the set of hyperparameters with the highest probability of optimisation according to Bayes' theorem. This speeds up the process of finding optimal hyperparameters compared to alternative search algorithms such as grid search or random search, and has been shown the find a better set of hyperparameters in the same number of runs \citep{bayes}. When hyperparameter tuning, the performance of each set of hyperparameters is averaged over 10 runs to mitigate the effect of fluctuations from different initialisations and data splits.

\subsection{Feature Calculation} \label{feature calculation}

In addition to these hyperparameters, there are other variables we can control. As well as considering which of these features to include, these features also depend on the threshold, $\nu$, in the case of MFs and EFDs, and order, $N$, in the case of EFDs. Therefore these two parameters were added to the list of hyperparameters which were varied during the tuning stage. The thresholds can vary in two ways: the number of thresholds the features are calculated from, and the values of each of these thresholds. We keep the number of thresholds as a hyperparameter and define the values of the thresholds to be linearly distributed from the minimum pixel value to the maximum pixel value for simplicity. 

Through imposing a threshold, the galaxy images are converted to binary images, from which the three MFs are calculated using \texttt{QuantImPy} \citep{quantimpy}. When testing the MFs, we choose to investigate not only the value of the MFs, but also how the MFs vary with threshold. This was motivated by the hypothesis that information of the FR-type could be contained in the relative values of features at different thresholds. For example, as the threshold increases, we might find that in FR\,I galaxies the regions above this threshold tend to form one connected region, while FR\,II galaxies might tend to form two connected regions. This difference would be quantified in MF 2, the Euler characteristic, as shown in Figure~\ref{fig:mfs}. For $t$ number of thresholds, this results in 3 $\times\:t$ different data points to feed into the model.  

HFs do not depend on threshold or order, and are calculated from an image of integer pixel values. As explained in Section~\ref{elliptical fourier descriptors}, we only calculate the first 13 of the 14 HFs, which is done using the \texttt{Mahotas} library \citep{mahotas}\footnote{In the original paper by \cite{haralick_og}, the equations for the HFs are given in the appendix. However, it appears that there is an error in the seventh feature, referred to as `sum variance', which is defined as the expected value of the squared deviation from the `sum entropy', as opposed to the mean as in any usual definition of variance. The code gives the option to ignore this error, which we employ.}. To allow for these features to be rotationally invariant, the 13 HFs are calculated four times, for each of the four ways two adjacent pixels can be related to each other (vertically adjacent, horizontally adjacent, and adjacent on both diagonals). \textcolor{black}{As we are interested in the difference between adjacent pixels, we use a distance offset of 1. This allows the Haralick features to capture information with the same resolution as the image.} These are averaged over to produce the set of 13 HFs which we use.

Similarly to MFs, the number of EFDs calculated depends on the number of thresholds, $t$, and also the order, $N$, to which the Fourier sum is expanded, with $t~\times~(4N-3)$ EFDs in total. At $N=1$, the Fourier analysis results in an ellipse with its major axis extending in the longest direction of the contour, which we align with the coordinate axes as described in Section~\ref{elliptical fourier descriptors}. This results in the coefficients, which we extract using the \texttt{PyEFD} library \citep{pyefd}, being rotationally invariant. 

\section{Model Performance}
\label{sec:performance}

\begin{figure*}
    \centering
    \includegraphics[width=\textwidth]{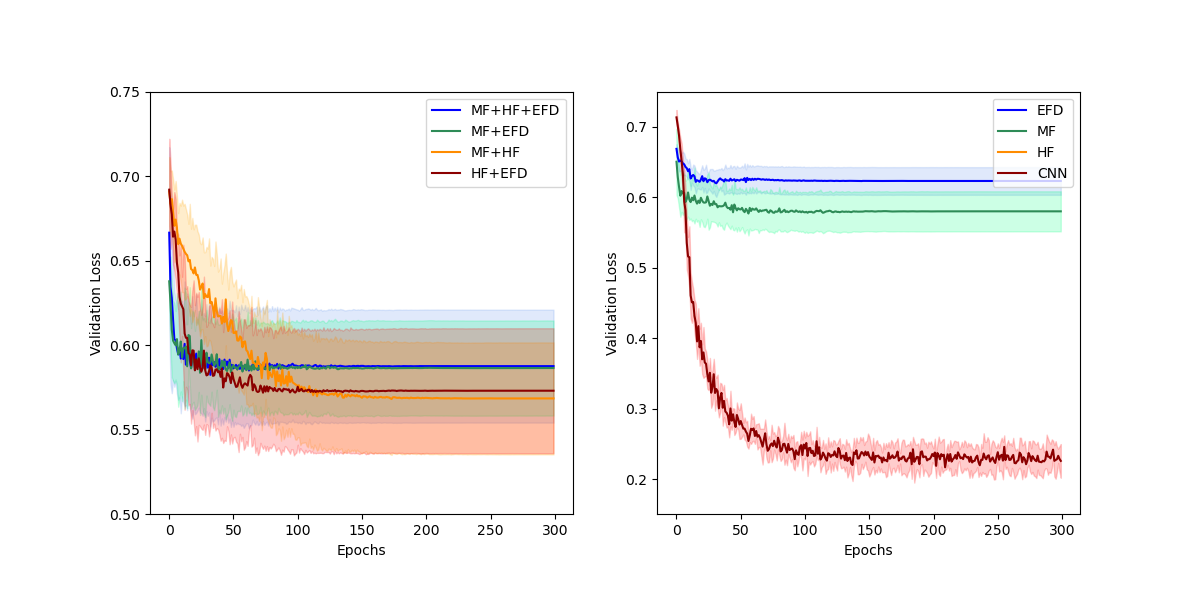}
    \caption{Left: Validation loss as a function of epoch for the neural network architecture described in Section~\protect\ref{sec:nn_method} for different combinations of equivariant features. Right: the validation loss for the CNN from \protect\cite{E2CNN} is shown for comparison.}
    \label{fig:valcurves}
\end{figure*}

When assessing the performance of the supervised learning models, we use the metrics of recall, precision, F1 score and accuracy, calculated against a reserved test set. 
Tables~\ref{tab:xgbmetrics}~and~\ref{tab:nnmetrics} give the values of these metrics for the XGBoost and neural network models respectively, for all combinations of the three types of features set. \textcolor{black}{These are tested using the best set of hyperparameters, averaged over 10 runs with different random initialisations from which the standard deviation is calculated. The computational times reported are the sums of the time taken for training and testing using an Apple M2 chip (CPU), including the time taken to calculate features. Furthermore, we include that training to validation loss ratio as an indicator of generalisability of the model.} All  performance metrics are above 50\%, demonstrating that these features contain information about the FR type of the galaxies. XGBoost consistently outperforms the neural network in every case, with a mean  difference in accuracy of $\sim$$5\%$. The uncertainty in accuracy and F1 score between the neural network and XGBoost runs are generally comparable, suggesting that although XGBoost may perform better, the methods are similar in terms of stability. When applying the neural network to MFs alone, we find a large variation in the recall of both FR\,I and FR\,II, indicating that similar accuracies can be achieved through different distributions of predictions. For FR\,I classification, the recall is higher than the precision for every run, while for FR\,II classification the opposite is the case, suggesting a disproportionate labelling of galaxies as FR\,II. In both machine learning models, the best feature set combination is MFs and HFs, producing an accuracy of $74.2\pm1.3\%$ for the neural network and $77.3\pm2.0\%$ for XGBoost. The validation loss over the number of epochs for all neural network runs is shown in Figure~\ref{fig:valcurves}.

\begin{table*}
\centering
\caption{Performance metrics for classification of the MiraBest dataset using XGBoost. Column [1] lists the type of features being used for classification in each case. Run times were obtained on a laptop with an Apple M2 chip (CPU). The best values for each metric are shown in bold.}
\label{tab:xgbmetrics}
\resizebox{\textwidth}{!}{%
\begin{tabular}{|c|c|ccc|ccc|c|}
\hline
\multirow{2}{*}{\textbf{Features}} & \multirow{2}{*}{\textbf{Accuracy [\%]}} & \multicolumn{3}{c|}{\textbf{FR\,I}}                                                                                            & \multicolumn{3}{c|}{\textbf{FR\,II}}                                                                                           & \multirow{2}{*}{\shortstack[l]{\textbf{Computation}\\ \textbf{Time [sec]}}} \\ \cline{3-8}
                                   &                                         & \multicolumn{1}{c|}{\textbf{Precision}}         & \multicolumn{1}{c|}{\textbf{Recall}}            & \textbf{F1-score}          & \multicolumn{1}{c|}{\textbf{Precision}}         & \multicolumn{1}{c|}{\textbf{Recall}}            & \textbf{F1-score}          &                                                                             \\ \hline
\textbf{MF}                        & 76.54 $\pm$ 2.20                        & \multicolumn{1}{c|}{0.708 $\pm$ 0.021}          & \multicolumn{1}{c|}{\textbf{0.855 $\pm$ 0.041}} & 0.774 $\pm$ 0.022          & \multicolumn{1}{c|}{\textbf{0.843 $\pm$ 0.035}} & \multicolumn{1}{c|}{0.685 $\pm$ 0.035}          & 0.755 $\pm$ 0.025          & 18.4 $\pm$ 0.5                                                              \\ \hline
\textbf{HF}                        & 70.19 $\pm$ 1.72                        & \multicolumn{1}{c|}{0.671 $\pm$ 0.019}          & \multicolumn{1}{c|}{0.722 $\pm$ 0.028}          & 0.695 $\pm$ 0.018          & \multicolumn{1}{c|}{0.735 $\pm$ 0.019}          & \multicolumn{1}{c|}{0.684 $\pm$ 0.028}          & 0.708 $\pm$ 0.019          & \textbf{12.4 $\pm$ 0.7}                                                     \\ \hline
\textbf{EFD}                       & 69.71 $\pm$ 2.44                        & \multicolumn{1}{c|}{0.668 $\pm$ 0.031}          & \multicolumn{1}{c|}{0.714 $\pm$ 0.032}          & 0.690 $\pm$ 0.023          & \multicolumn{1}{c|}{0.728 $\pm$ 0.022}          & \multicolumn{1}{c|}{0.682 $\pm$ 0.044}          & 0.704 $\pm$ 0.029          & 114.3 $\pm$ 0.9                                                             \\ \hline
\textbf{MF+HF}                     & \textbf{77.31 $\pm$ 1.98}               & \multicolumn{1}{c|}{0.724 $\pm$ 0.025}          & \multicolumn{1}{c|}{0.841 $\pm$ 0.024}          & \textbf{0.777 $\pm$ 0.018} & \multicolumn{1}{c|}{0.834 $\pm$ 0.020}          & \multicolumn{1}{c|}{0.713 $\pm$ 0.036}          & \textbf{0.768 $\pm$ 0.024} & 17.2 $\pm$ 0.4                                                              \\ \hline
\textbf{MF+EFD}                    & 76.25 $\pm$ 2.92                        & \multicolumn{1}{c|}{\textbf{0.725 $\pm$ 0.033}} & \multicolumn{1}{c|}{0.802 $\pm$ 0.040}          & 0.761 $\pm$ 0.029          & \multicolumn{1}{c|}{0.806 $\pm$ 0.033}          & \multicolumn{1}{c|}{0.727 $\pm$ 0.042}          & 0.764 $\pm$ 0.031          & 68.1 $\pm$ 0.8                                                              \\ \hline
\textbf{HF+EFD}                    & 74.71 $\pm$ 3.77                        & \multicolumn{1}{c|}{0.711 $\pm$ 0.044}          & \multicolumn{1}{c|}{0.784 $\pm$ 0.036}          & 0.745 $\pm$ 0.035          & \multicolumn{1}{c|}{0.787 $\pm$ 0.033}          & \multicolumn{1}{c|}{0.715 $\pm$ 0.055}          & 0.749 $\pm$ 0.041          & 40.9 $\pm$ 1.7                                                              \\ \hline
\textbf{MF+HF+EFD}                 & 75.38 $\pm$ 2.66                        & \multicolumn{1}{c|}{0.721 $\pm$ 0.029}          & \multicolumn{1}{c|}{0.780 $\pm$ 0.034}          & 0.749 $\pm$ 0.027          & \multicolumn{1}{c|}{0.789 $\pm$ 0.029}          & \multicolumn{1}{c|}{\textbf{0.731 $\pm$ 0.034}} & 0.758 $\pm$ 0.027          & 14.8 $\pm$ 0.4                                                              \\ \hline
\end{tabular}%
}
\end{table*}

\begin{table*}
\centering
\caption{Performance metrics for classification of the MiraBest dataset using the neural network described in Section~\protect\ref{sec:nn_method}. The CNN metrics are taken from \protect\cite{E2CNN}. Column [1] lists the type of features being used for classification in each case. Run times were obtained on a laptop with an Apple M2 chip (CPU). The best values for each metric are shown in bold.}
\label{tab:nnmetrics}
\resizebox{\textwidth}{!}{%
\begin{tabular}{|c|c|ccc|ccc|c|l|}
\hline
\multirow{2}{*}{\textbf{Features}} & \multirow{2}{*}{\textbf{Accuracy [\%]}} & \multicolumn{3}{c|}{\textbf{FR\,I}}                                                                                            & \multicolumn{3}{c|}{\textbf{FR\,II}}                                                                                           & \multirow{2}{*}{\shortstack[l]{\textbf{Computation}\\ \textbf{Time [sec]}}} & \multirow{2}{*}{\shortstack[l]{\textcolor{black}{\textbf{Validation-Training}}\\ \textcolor{black}{\textbf{Loss Ratio}}}} \\ \cline{3-8}
                                   &                                         & \multicolumn{1}{c|}{\textbf{Precision}}         & \multicolumn{1}{c|}{\textbf{Recall}}            & \textbf{F1-score}          & \multicolumn{1}{c|}{\textbf{Precision}}         & \multicolumn{1}{c|}{\textbf{Recall}}            & \textbf{F1-score}          &                                                                             &                                                                                     \\ \hline
\textbf{MF}                        & 67.98 $\pm$ 1.43                        & \multicolumn{1}{c|}{0.643 $\pm$ 0.025}          & \multicolumn{1}{c|}{0.735 $\pm$ 0.098}          & 0.681 $\pm$ 0.037          & \multicolumn{1}{c|}{0.736 $\pm$ 0.049}          & \multicolumn{1}{c|}{0.631 $\pm$ 0.082}          & 0.673 $\pm$ 0.031          & 32.2 $\pm$ 0.6                                                              & \textcolor{black}{0.974 $\pm$ 0.057}                                                                   \\ \hline
\textbf{HF}                        & 67.50 $\pm$ 1.54                        & \multicolumn{1}{c|}{0.635 $\pm$ 0.024}          & \multicolumn{1}{c|}{0.739 $\pm$ 0.040}          & 0.682 $\pm$ 0.011          & \multicolumn{1}{c|}{0.635 $\pm$ 0.024}          & \multicolumn{1}{c|}{0.618 $\pm$ 0.056}          & 0.667 $\pm$ 0.030          & 43.5 $\pm$ 1.2                                                              & \textcolor{black}{0.945 $\pm$ 0.062}                                                                   \\ \hline
\textbf{EFD}                       & 63.75 $\pm$ 3.04                        & \multicolumn{1}{c|}{0.610 $\pm$ 0.035}          & \multicolumn{1}{c|}{0.651 $\pm$ 0.069}          & 0.627 $\pm$ 0.038          & \multicolumn{1}{c|}{0.670 $\pm$ 0.031}          & \multicolumn{1}{c|}{0.625 $\pm$ 0.072}          & 0.644 $\pm$ 0.042          & 85.9 $\pm$ 1.3                                                              & \textcolor{black}{0.657 $\pm$ 0.023}                                                                   \\ \hline
\textbf{MF+HF}                     & \textbf{74.23 $\pm$ 1.28}               & \multicolumn{1}{c|}{\textbf{0.697 $\pm$ 0.013}} & \multicolumn{1}{c|}{\textbf{0.804 $\pm$ 0.046}} & \textbf{0.746 $\pm$ 0.020} & \multicolumn{1}{c|}{\textbf{0.800 $\pm$ 0.030}} & \multicolumn{1}{c|}{\textbf{0.687 $\pm$ 0.029}} & \textbf{0.738 $\pm$ 0.011} & \textbf{21.8 $\pm$ 1.2}                                                     & \textcolor{black}{0.970 $\pm$ 0.061}                                                                   \\ \hline
\textbf{MF+EFD}                    & 67.50 $\pm$ 2.61                        & \multicolumn{1}{c|}{0.638 $\pm$ 0.021}          & \multicolumn{1}{c|}{0.718 $\pm$ 0.066}          & 0.674 $\pm$ 0.036          & \multicolumn{1}{c|}{0.720 $\pm$ 0.041}          & \multicolumn{1}{c|}{0.636 $\pm$ 0.038}          & 0.674 $\pm$ 0.023          & 89.0 $\pm$ 1.5                                                              & \textcolor{black}{0.950 $\pm$ 0.045}                                                                   \\ \hline
\textbf{HF+EFD}                    & 71.25 $\pm$ 2.41                        & \multicolumn{1}{c|}{0.668 $\pm$ 0.032}          & \multicolumn{1}{c|}{0.780 $\pm$ 0.037}          & 0.719 $\pm$ 0.024          & \multicolumn{1}{c|}{0.769 $\pm$ 0.029}          & \multicolumn{1}{c|}{0.653 $\pm$ 0.050}          & 0.705 $\pm$ 0.033          & 64.3 $\pm$ 0.9                                                              & \textcolor{black}{0.793 $\pm$ 0.048}                                                                   \\ \hline
\textbf{MF+HF+EFD}                 & 67.69 $\pm$ 1.37                        & \multicolumn{1}{c|}{0.635 $\pm$ 0.015}          & \multicolumn{1}{c|}{0.743 $\pm$ 0.042}          & 0.684 $\pm$ 0.018          & \multicolumn{1}{c|}{0.731 $\pm$ 0.024}          & \multicolumn{1}{c|}{0.618 $\pm$ 0.037}          & 0.669 $\pm$ 0.019          & 39.7 $\pm$ 0.6                                                              & \textcolor{black}{\textbf{0.979 $\pm$ 0.055}}                                                                   \\ \hline
\hline
\textbf{CNN}                       & 94.04 $\pm$ 1.37                        & \multicolumn{1}{c|}{0.935 $\pm$ 0.018}          & \multicolumn{1}{c|}{0.940 $\pm$ 0.024}          & 0.937 $\pm$ 0.015          & \multicolumn{1}{c|}{0.946 $\pm$ 0.015}          & \multicolumn{1}{c|}{0.941 $\pm$ 0.018}          & 0.944 $\pm$ 0.013          & 2410.2 $\pm$ 7.7                                                            & \textcolor{black}{0.917 $\pm$ 0.068}                                                                   \\ \hline
\end{tabular}%
}
\end{table*}

\section{Discussion}
\label{sec:discussion}

When tested on their own, MFs outperform HFs and EFDs, with an accuracy of $76.5\pm2.2\%$ compared to $70.2\pm1.7\%$ and $69.7\pm2.4\%$ using XGBoost, respectively. Combining EFDs and HFs produces the biggest increase in the individual performances, resulting in an accuracy of $74.7\pm3.8\%$. However, combining EFDs and MFs using XGBoost results in a slightly worse accuracy than the MFs on their own, suggesting a significant overlap in the information that these features contain, and also an inability for these models to ignore data that is superfluous. Combining MFs and HFs produces the model with the best accuracy and F1 scores for both the neural network and XGBoost, with the XGBoost algorithm using these features having the highest overall accuracy. Compared to using XGBoost with MFs alone, adding HFs increases the FR\,II recall and FR\,I precision, but at the expense of a decrease in FR\,I recall and FR\,II precision, suggesting an increase in the number of galaxies being labelled as FR\,I. Although this model has the best performance, the accuracies of all the XGBoost runs that combine two or more feature types lie within their common uncertainty bounds, while there is a larger variation in performance using the neural network. 

In the study by \cite{E2CNN}, the accuracy of a non-equivariant CNN is found to be $94.0\pm1.4\%$, and the accuracy of a $\rm D_{16}$-equivariant CNN is $96.6\pm1.3\%$, which our models are far from achieving. However, the computational complexity of convolutional layers means that the models developed in this work are significantly faster, as shown in Table~\ref{tab:nnmetrics}. Using a CNN results in near 50 times the required computation time  of the non-convolutional machine learning algorithms, meaning these can provide a significantly faster classification of large amounts of data when a high accuracy is not required. \textcolor{black}{Furthermore, as shown in Table \ref{tab:nnmetrics}, all feature sets besides EFD and HF+EFD have similar validation-training loss ratios to that of the CNN, indicating that there is no particular advantage in either approach with respect to generalisation.} Fluctuations in computation times for the models developed in this study are primarily due to the different order and number of thresholds from which the features were calculated, as these hyperparameters were tuned individually for each model. 

\textcolor{black}{We note that the improved performance of the CNN compared with the direct feature extraction methods is not expected to be robust to reductions in training data volume \citep[see e.g.][]{cnn-vs-glcm}, as CNNs require relatively large data volumes to train effectively. }

\subsection{Feature Importance}
\label{sec:importance}

\begin{figure}
    \centering
    \includegraphics[width=0.5\textwidth]{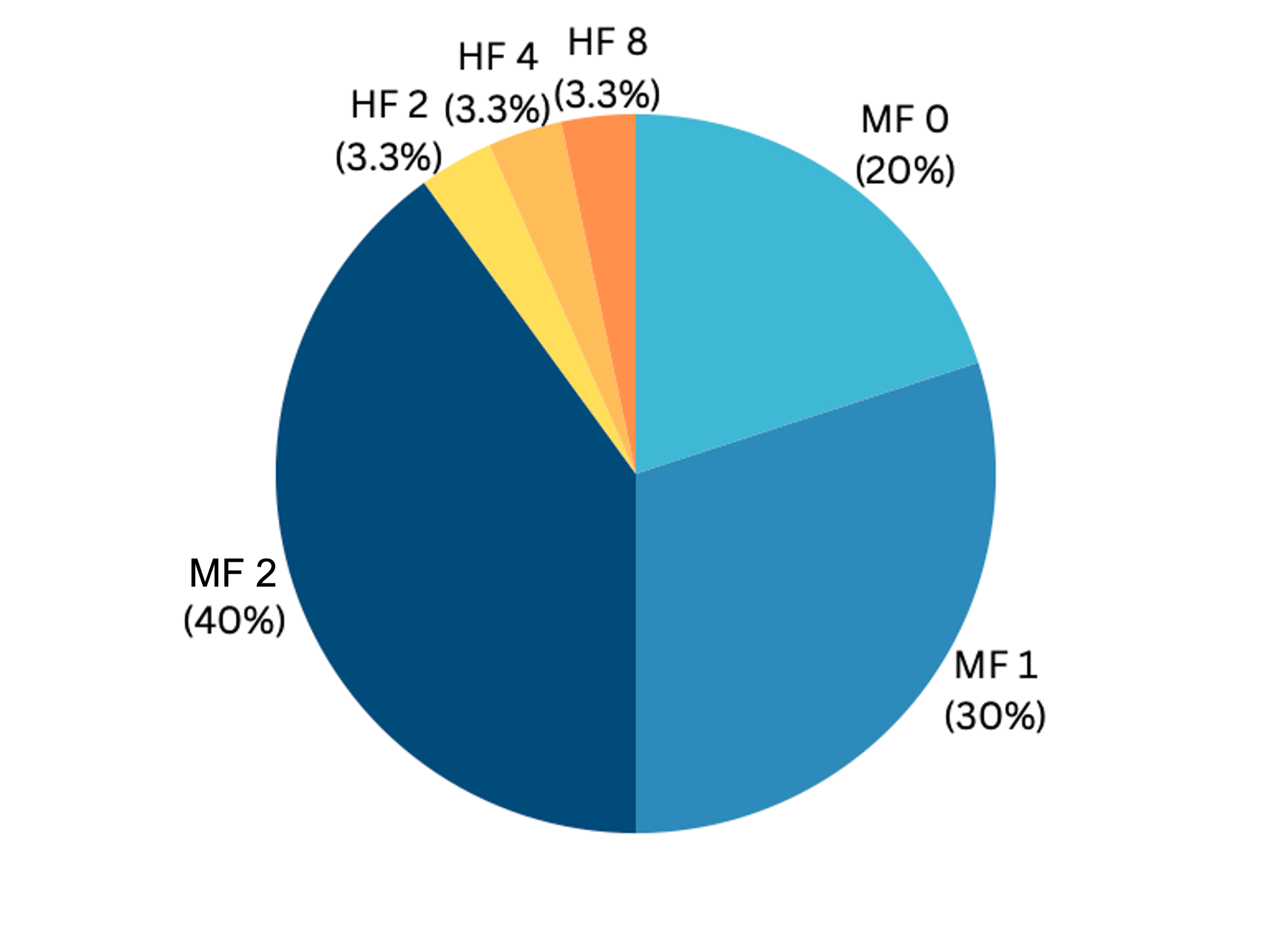}
    \caption{The relative importance of each type of feature for the top 30 most important features from the best run according to XGBoost.}
    \label{fig:feature importance}
\end{figure}

In addition to the performance of the models trained on the individual features, we can investigate their suitability for FR classification through inspecting their relative importance according to XGBoost. Shown in Figure~\ref{fig:feature importance} are the most important features according to XGBoost from the run of highest accuracy (MFs+HFs). MFs make up the majority of these most important features, with 40\% of these being the MF corresponding to the Euler characteristic evaluated at different thresholds. Of the 13 HFs, the most important were HF\,8 (sum entropy), HF\,2 (contrast) and HF\,4 (sum of squares variance), listed in decreasing importance. 

During the model tuning process, the number of thresholds and order of EFDs are varied as hyperparameters. Order is found to be very weakly correlated with the performance of the model, and when EFDs were combined with any other feature, fewer orders were favoured. The number of thresholds was also found to be only weakly correlated to the performance in comparison with other hyperparameters. Having very few thresholds results in generally a worse performance, but beyond $\sim$5 thresholds, adding more did not increase the accuracy. This is unsurprising as the thresholds always span the entirety of the possible values, and the cumulative nature of increasing the cutoff value for the pixels means that adding more thresholds does not provide any new information beyond increasing the resolution at which the change in the values of the MFs and EFDs can be calculated. \textcolor{black}{The only Haralick feature that is not robust to noise is the correlation \cite[HF 3; see e.g.][]{Brynolfsson:2017aa}. Our results show that for our best performing model, which uses a combination of the Haralick features and Minkowski functionals, HF 3 (Equation~\ref{eq:correlation}) is not in the top 30 most important features, as can be seen in Figure~\ref{fig:feature importance}. However, when using Haralick features on their own, this feature ranks in the top three most important and therefore a model based on Haralick features alone is expected to be less robust to noise.}

\subsection{Representation Space}
\label{sec:representations}

To investigate the distribution of features in more detail, we use the Uniform Manifold Approximation and Projection (UMAP) algorithm, which is a non-linear dimensional reduction technique developed by \cite{umap}. The aim of UMAP is to project high dimensional data into a lower dimensional space (typically two dimensions to allow for visualisation), while preserving topological relationships between the datapoints. This is not a supervised machine learning algorithm and does not aim to classify the data according to the labels given, but instead groups the data based on how close they are in parameter space. We can gain an understanding of how the features intrinsically contain information about the galaxy type by investigating what galaxies are mapped to similar areas of the manifold. This allows a comparison between the features, and allows us to analyse the suitability of these features in the context of radio galaxy FR classification, especially in the context of neural networks which are typical hard to interpret. UMAP has been found to provide a useful tool for visualising the separation of galaxies from stars and quasars \citep{anna_umap}, and performs better when working on large scales than t-distributed Stochastic Neighbour Embedding (t-SNE), another tool commonly used in dimensionality reduction \citep{umap}. 

Input data to the UMAP algorithm, which typically is not uniformly distributed in the high dimensional parameter space, is mapped to a Reimannian manifold in which the data is uniformly distributed by construction. Reimannian geometry tells us that a hypersphere of unit radius on this manifold maps to a hypersphere that stretches to the $k^{\rm th}$ nearest neighbour according to a chosen metric $d$ in the parameter space. The value of $k$ is a hyperparameter which is used to define a notion of locality, and results in each datapoint having its own sized hypersphere of locality in the parameter space. 
Two datapoints are assigned a probability of being connected based on the overlap of these hyperspheres, and these probabilities are used as the weights of the edges of a connected graph with datapoints as nodes. This results in a topological representation of the data which has been constructed to preserve the topology of the original manifold, and is easier to manipulate onto a new lower dimensional manifold. This lower dimensional manifold is constructed by creating a generic lower dimensional manifold and then manipulating it to optimise its information overlap with that of the connected graph, quantified by the cross entropy. The optimisation process is done using a force directed graph layout algorithm, in which datapoints in the same clump are attracted and datapoints in different clumps are repelled until an optimal representation is reached.

\begin{figure*}
    \centering
    \includegraphics[width=\textwidth]{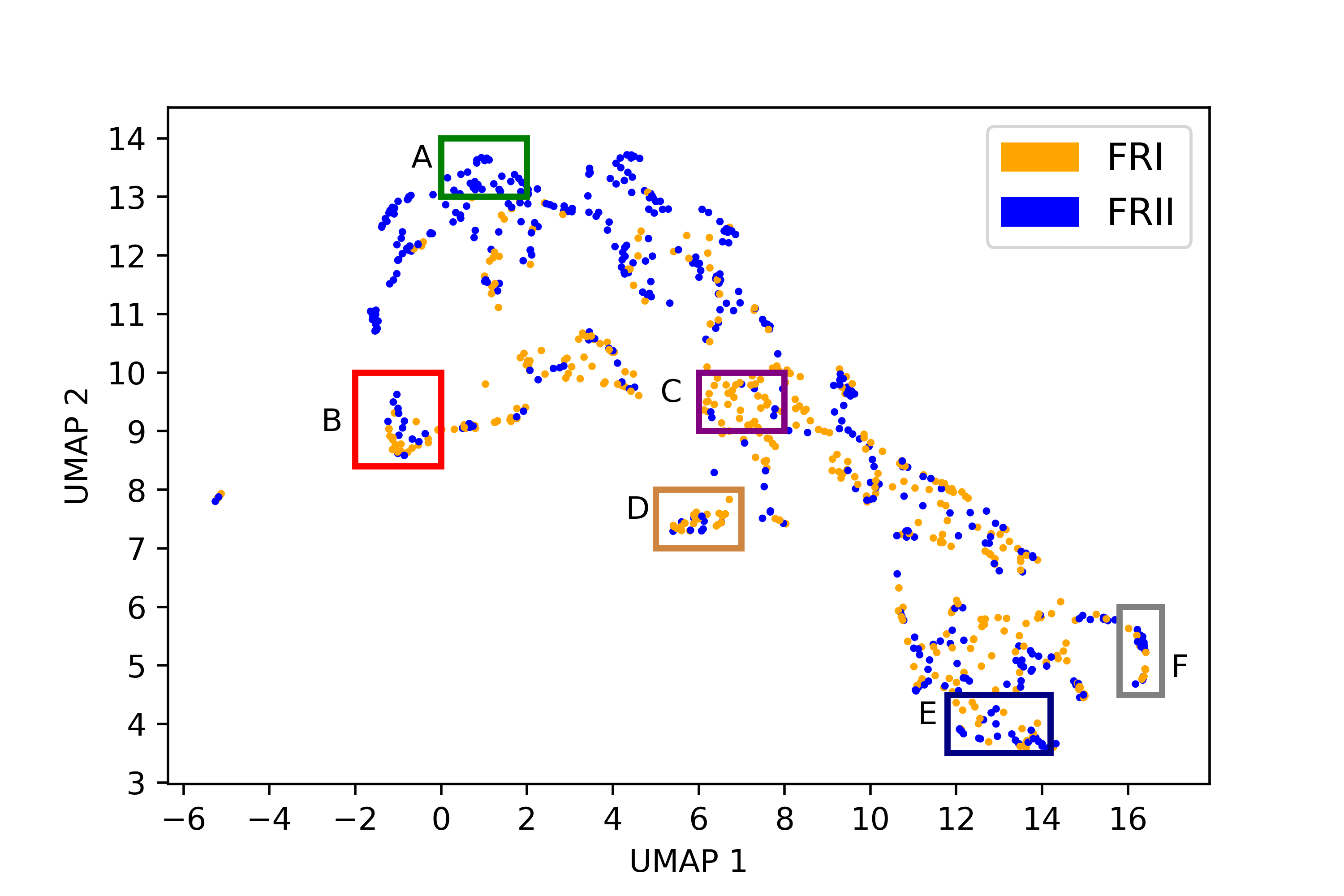}
    \caption{UMAP embedding of the top 10 features from the most accurate XGBoost run. Labels of FR\,I and FR\,II types are given in orange and blue respectively, with labelled regions of interest. A Euclidean metric is used with 8 nearest neighbours and minimum distance of 0.1.}
    \label{fig:umap}
\end{figure*}

\begin{figure*}
    \centering
    \includegraphics[width=\textwidth]{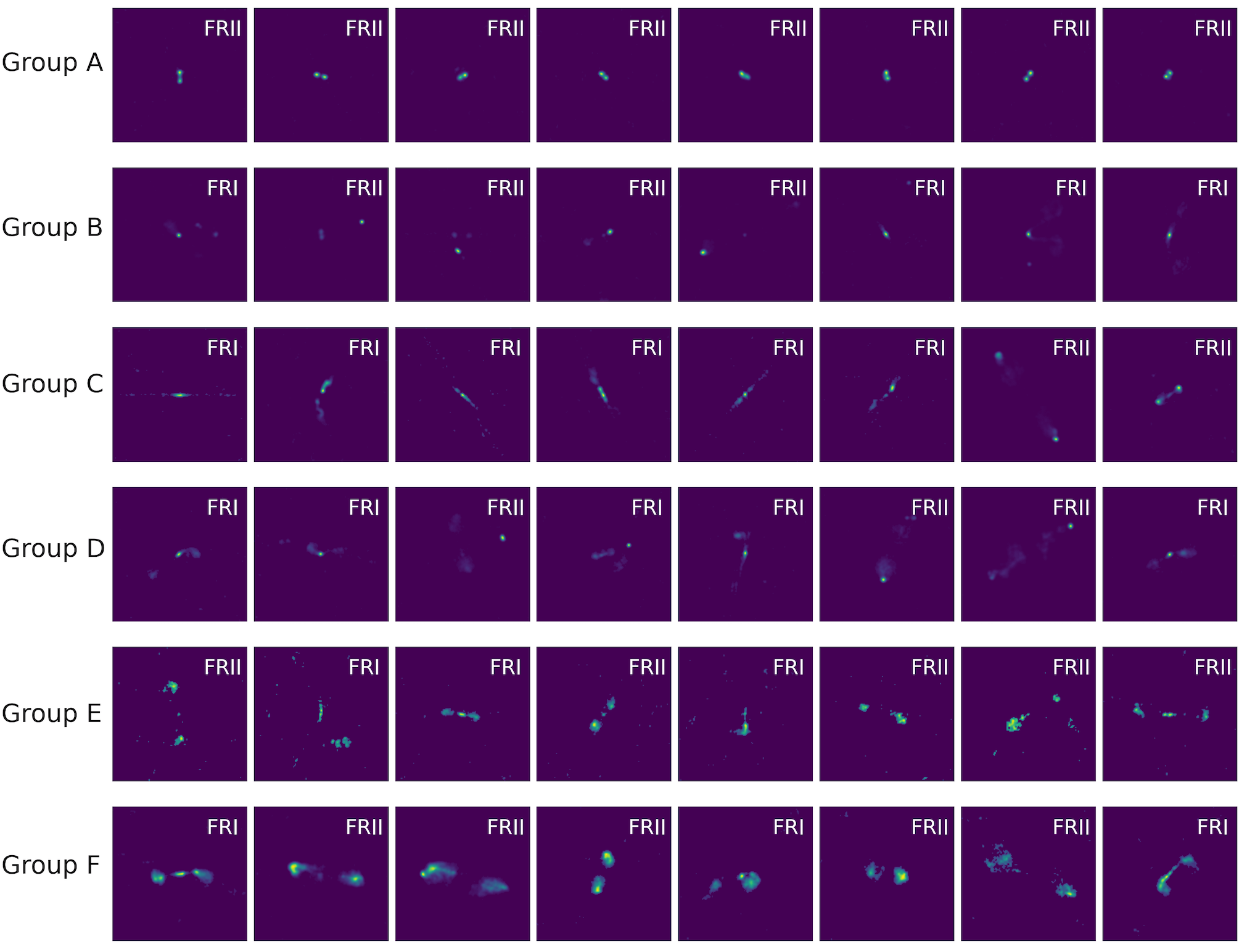}
    \caption{Random galaxies from each of the 6 regions labelled in Figure~\ref{fig:umap}. FR type for each galaxy is labelled.}
    \label{fig:groups}
\end{figure*}

We construct a UMAP projection of the equivariant feature space in two dimensions using the top 10 most important features from the feature importance analysis in Section~\ref{sec:importance}, which is presented in Fig~\ref{fig:umap} with the true labels of the galaxies indicated. There is no distinct separation between the two types into groups, although there are regions of denser FR\,I (near the area marked C) and FR\,II (near the region marked A). The lack of obvious clustering of FR\,I and FR\,II is to be expected from the relatively low accuracy of the supervised machine learning models, but the distinction between the two types in certain areas is of interest. Examples of the galaxies in each of the 6 regions of Figure~\ref{fig:umap} is shown in Figure~\ref{fig:groups}.

The largest distance between regions in Figure~\ref{fig:umap} is that between group A and groups E and F. As shown in Figure~\ref{fig:groups}, going from group A to group E corresponds to the galaxies becoming more extended and containing a higher total image flux. UMAP tends to group datapoints that are close in the parameter space and spread out the datapoints that are further away, and the fact that this transition from compact sources to extended sources aligns with the principle axis of the embedding suggests that this information is the primary information held in these features. 

Another feature of the galaxies that appears to be represented in the embedding is that of messiness of the image. Galaxies in groups A, C and F, on the upper side of the UMAP embedding, tend to be more organised and show more homogeneity within the group. Group A consists exclusively of compact sources with two distinct areas of luminosity and are almost entirely FR\,II. Group C consists of sources that are thinly extended in one direction, most of which are FR\,I.  Group F has a more even mix of FR\,I and FR\,II, but they are consistent in that they contain two large bright regions that are well separated, though may contain a third bright but compact region at the centre of the galaxy. In comparison, groups B, D and E of the lower side of the UMAP embedding typically contain more distinct regions, many of which are less well-defined. Group B contains compact bright sources that are separated from compact dimmer regions. The large empty regions between the multiple compact areas of brightness make it hard to tell where the centre of the galaxy is and it often is not clear if these areas of brightness are from the same source. Group D and group E  both contain galaxies with multiple components that tend to be less straight than those in group C, though the galaxies in group D tend to have one brighter region and multiple dimmer regions, while galaxies in group C are more uniformly bright and display high contrast. 

\subsection{Restriction of Invariance}
\label{sec:constriction}

Both MFs and HFs are E$(2)$ invariant. As invariant features are a subset of equivariant features, there are inherent limits to the information they contain about galaxy morphology. Despite the E$(2)$ equivariance of EFDs, challenges arose in the presence of multiple contours. Analysis of Equation \ref{sumcoeffs} reveals a potential issue with concatenating multiple contours as the gradient is undefined at the point of concatenation. This effect limits the functionality of the elliptical Fourier analysis and leads to a loss of information about the distances between contours.

Analysis of Figures~\ref{fig:umap} and \ref{fig:groups} reveals that MFs and HFs inherently capture information about galaxy extension. This information about extension is built into the MFs, as both area and perimeter are measures of size. How HFs sort by extension is more enigmatic, as they measure the relationships between the pixels in an image, regardless of whether or not they are part of the background. Therefore, the normalisation constant from Equation \ref{normharalick} is always the same for all galaxies, and as such HFs also encode information about galaxy extension. It is important to note that while total extension is a useful metric to classify FR galaxies, its dependence on telescope resolution and cosmological distance introduces a caveat. The accuracy of  models based on feature sets such as those presented here when applied to new data at different telescope resolutions will be affected by this dependence.

\textcolor{black}{We suggest that the limited performance of the E(2) equivariant features is due to the nature of the features being invariant rather than equivariant to the symmetry of the problem and therefore limiting their flexibility in the context of FR classification. These features have demonstrated their strength as useful tools for classification in less complex image-processing tasks when compared to CNNs, such as in medical imaging \citep{cnn-vs-glcm}, indicating that these features do provide viable alternatives to CNNs in some contexts. However, as the task of FR classification for radio galaxies is inherently complex due to the visual variation found in each FR type, these features alone are unable to extract all the information from the image that is relevant to their FR classification, explaining their limited performance in the context of this research.}

\section{Conclusions}
\label{sec:conclusion}

In this work we have compared MFs, HFs and EFDs as tools for the task of FR classification of radio galaxies. These features are equivariant under E(2) symmetries, a desirable property for this task. When used as inputs to both a neural network and XGBoost, we find that all three of these feature sets contain morphological information about the galaxies, with MFs being most informative and EFDs least informative. We demonstrate that although combining these features can result in an increase in performance, this increase is only incremental due to the limitations of these features when applied to complex real-world data, and information overlap between the different types of features. 

We show that, compared to equivalent group-equivariant CNNs, the models developed in this work are around 50 times faster at classifying radio galaxies, though their accuracy is found to be limited to $\sim80\%$. 

Similarly to \cite{haralick_radio}, our analysis shows that these features primarily capture morphological information such as extension and entropy, which are only weakly correlated to FR type. Therefore we suggest that these features may be more applicable for the task of unsupervised clustering of data where higher computational efficiency is required. 

\section*{Acknowledgements}

AMS gratefully acknowledges support from the UK Alan Turing Institute under grant reference EP/V030302/1.

\section*{Data Availability}

This work makes use of the MiraBest machine learning dataset, which is publically available under a Creative Commons 4.0 license at \url{https://doi.org/10.5281/zenodo.4288837}.



\bibliographystyle{mnras}
\bibliography{mf_paper}


\begin{onecolumn}
\appendix

\section{Equivariant Features} \label{app:featuretheory}

\subsection{Minkowski Functionals}
\label{app:mf}

%
%
%
%
From the imposed threshold $\nu$, the excursion set $\Sigma(\nu)$ is defined as the set of points above this threshold. This threshold converts the input image to a binary form, where $\partial \Sigma(\nu)$ is the boundary of the excursion set and $\kappa$ is the curvature of the boundary. In two dimensions, the three MFs are
\begin{equation} \label{Minkfuncs}
    \begin{split}
    MF_0(\nu) & = \int_{\Sigma (\nu)} \mathrm{d}A, \\
    MF_1(\nu) & = \frac{1}{2\pi}\int_{\partial \Sigma (\nu)} \mathrm{d}S, \\
    MF_2(\nu) & = \frac{1}{2\pi^{2}}\int_{\partial \Sigma (\nu)}\kappa \, \mathrm{d}S,
    \end{split}
\end{equation}
\citep{Parroni:2020}.
%
They can be related to the surface area, $A=MF_{0}$, the perimeter, $C=2\pi MF_{1}$, and the Euler characteristic, $\chi = \pi MF_{2}$, which is the number of disconnected areas in an object minus the number of disconnected areas outside the object. 
%
%

\subsection{Haralick Features} \label{app:haralicktheory}

\begin{figure*}
    \centering
    \includegraphics[scale=0.8]{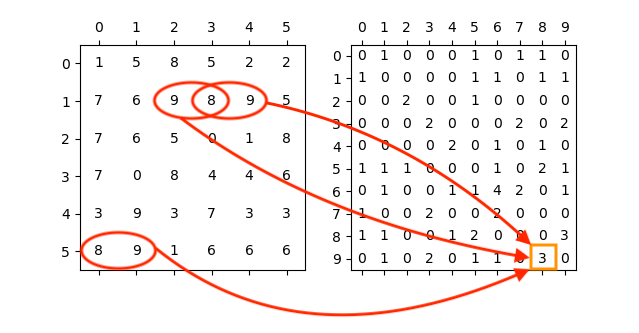}
    \caption{Co-occurrence matrix (right) calculated for a random matrix (left) according to a horizontal relation between pixels. In red, we show the calculation for the $i=8$ and $j=9$ entry. As this relation is symmetric, the co-occurrence matrix is also symmetric.}
    \label{coocurrenceexmpl}
\end{figure*}

There are 14 HFs, but as the last feature is considered computationally unstable, only the first 13 are commonly used \citep{Brynolfsson:2019}. They are calculated through the co-occurrence matrix, ${P}_{\theta}$, defined as
\begin{equation} \label{COM}
P_{\theta}(i, j) = \sum_{n_{x}=1}^{N_{x}}\sum_{n_{y}=1}^{N_{y}} f_{ij}(n_x,n_y),
\end{equation}
where
\begin{equation}
f_{ij}(n_x,n_y)=\begin{cases}
1,\; \quad $if$\; I(n_{x}, n_{x})=i\; $and$\; I(n_{x} + \Delta x, n_{x} + \Delta y)=j\\
0,\; \quad \rm{otherwise}
\end{cases}
\end{equation}
and $N_{x} \times N_{y}$ is the size of the image, $i$ and $j$ are possible intensity values, $\Delta x$ and $\Delta y$ are the distance offsets and $\theta \in \{ 0^\circ, 45^\circ, 90^\circ, 135^\circ \}$ is the direction of the offset between pixels. An example of how elements of a co-occurrence matrix is calculated is shown in Figure~\ref{coocurrenceexmpl}. Equation \ref{COM} is normalised into a probability matrix $p_{\theta}(i, j)=\frac{P_{\theta}(i, j)}{R}$ by the constant
\begin{equation} \label{normharalick}
R = \sum_{i}\sum_{j} P_{\theta}(i, j).
\end{equation}
When the features are calculated in all four directions and then averaged, rotational invariance is imposed.

\noindent
The equations for all 14 HFs, as defined in \cite{haralick_og} with the exception of the corrected version of HF 7, are as follows:
\begin{equation}
    \text{Angular second momentum: \hspace{1.5em}} HF_1=\sum_i \sum_j\left(p_{\theta} (i, j)\right)^2
\end{equation}

\begin{equation}
    \text{Contrast: \hspace{1.5em}} HF_2=\sum_{n=0}^{N_g-1} n^2\left(\sum_{\substack{i=1 \\|i-j|=n}}^{N_g} \sum_{\substack{j=1}}^{N_g} p_{\theta} (i, j)\right)
\end{equation}

\begin{equation} \label{eq:correlation}
    \text{Correlation: \hspace{1.5em}} HF_3=\frac{\sum_i \sum_j(i j) p_{\theta} (i, j)-\mu_x \mu_y}{\sigma_x \sigma_y}
\end{equation}

\begin{equation}
    \text{Sum of Squares Variance: \hspace{1.5em}} HF_4=\sum_i \sum_j(i-\mu)^2 p_{\theta} (i, j)
\end{equation}

\begin{equation}
    \text{Inverse Difference Moment: \hspace{1.5em}} HF_5=\sum_i \sum_j \frac{1}{1+(i-j)^2} p_{\theta} (i, j)
\end{equation}

\begin{equation}
    \text{Sum Average: \hspace{1.5em}} HF_6=\sum_{i=2}^{2 N_g} i p_{x+y}(i)
\end{equation}

\begin{equation}
    \text{Sum Variance (original): \hspace{1.5em}} HF_7=\sum_{i=2}^{2 N_g}\left(i-HF_8\right)^2 p_{x+y}(i)
\end{equation}
\begin{equation}
    \text{Sum Variance (corrected): \hspace{1.5em}} HF_7=\sum_{i=2}^{2 N_g}\left(i-HF_6\right)^2 p_{x+y}(i)
\end{equation}

\begin{equation}
    \text{Sum Entropy: \hspace{1.5em}} HF_8=-\sum_{i=2}^{2 N_g} p_{x+y}(i) \log \left(p_{x+y}(i)\right)
\end{equation}

\begin{equation}
    \text{Entropy: \hspace{1.5em}} HF_9=-\sum_i \sum_j p_{\theta} (i, j) \log (p_{\theta} (i, j))
\end{equation}

\begin{equation}
    \text{Difference Variance: \hspace{1.5em}} HF_{10}=\text { variance of } p_{x-y}
\end{equation}

\begin{equation}
    \text{Difference Entropy: \hspace{1.5em}} HF_{11}=-\sum_{i=0}^{N_{g-1}} p_{x-y}(i) \log \left(p_{x-y}(i)\right)
\end{equation}

\begin{equation}
    \text{Information Measure of Correlation A: \hspace{1.5em}} HF_{12}=\frac{H X Y-H X Y 1}{\max \{H X, H Y\}}
\end{equation}

\begin{equation}
    \text{Information Measure of Correlation B: \hspace{1.5em}} HF_{13}=\sqrt{1-\exp [-2(H X Y 2-H X Y)]}
\end{equation}

\begin{equation}
\text{Maximal Correlation Coefficient: \hspace{1.0em}} HF_{14}=\sqrt{\text {Second largest eigenvalue of } \sum_k \frac{p_{\theta}(i, k) p_{\theta}(j, k)}{p_x(i) p_y(k)}},
\end{equation}
where the notation is defined as:
\begin{equation}
    HX = -\sum_{i} p_{x}(i) \mathrm{log}(p_{x}(i)), \qquad
\end{equation}
\begin{equation}
    HY = -\sum_{j} p_{y}(j) \mathrm{log}(p_{y}(j))
\end{equation}
\begin{equation} H X Y 1=-\sum_i \sum_j p_{\theta} (i, j) \log \left(p_x(i) p_y(j)\right)\end{equation}
\begin{equation}H X Y 2=-\sum_i \sum_j p_x(i) p_y(j) \log \left(p_x(i) p_y(j)\right)\end{equation}
\begin{equation}H X Y=-\sum_i \sum_j p_{\theta} (i, j) \log (p_{\theta} (i, j))\end{equation}
\begin{equation}p_x(i)=\sum_{j=1}^{N_g} p_{\theta} (i, j)\end{equation}
\begin{equation}p_y(j)=\sum_{i=1}^{N_g} p_{\theta} (i, j)\end{equation}
\begin{equation}p_{x+y}(k)=\sum_{\substack{i=1 \\ i+j=k}}^{N_g} \sum_{j=1}^{N_g} p_{\theta} (i, j), \quad k=2,3,\dots 2N_g\end{equation}
\begin{equation}p_{x-y}(k)=\sum_{\substack{i=1 \\ |i-j|=k}}^{N_g} \sum_{j=1}^{N_g} p_{\theta} (i, j), \quad k=0,1,\dots N_g-1\end{equation}
\begin{equation}N_g= \text{Number of distinct grey levels in the quantised image.}\end{equation}

\subsection{Elliptical Fourier Descriptors} \label{app:efdtheory}

The Elliptical Fourier series computed to the $N^{{\rm th}}$ order is defined as 

\begin{equation} \label{fourierseries}
    \begin{bmatrix}
    x(t)\\ y(t)
    \end{bmatrix} \approx
    \begin{bmatrix}
    A_{0} \\ C_{0}
    \end{bmatrix} +
    \sum_{n=1}^{N} \begin{bmatrix}
    A_{n} & b_n\\
    C_{n} & d_n
    \end{bmatrix}
    \begin{bmatrix}
    \cos\left({\frac{2n\pi t}{T}}\right)\\
    \sin\left({\frac{2n\pi t}{T}}\right)
    \end{bmatrix}, \quad t\in(0, T],
\end{equation}
where $T$ is the period of the series, and the coefficients are calculated through
\begin{equation} \label{intcoeffs}
    \begin{split}
    \begin{bmatrix}
    A_{0}\\ C_{0}
    \end{bmatrix} & =
    \frac{1}{T} \int_{0}^{T}
    \begin{bmatrix}
    x(t)\\ y(t)
    \end{bmatrix} \mathrm{d}t, \\
    \begin{bmatrix}
    A_{n}\\ C_{n}
    \end{bmatrix} & =
    \frac{2}{T} \int_{0}^{T}
    \begin{bmatrix}
    x(t)\\ y(t)
    \end{bmatrix} \cos\left({\frac{2n\pi t}{T}} \right)\mathrm{d}t, \\
    \begin{bmatrix}
    B_{n}\\ D_{n}
    \end{bmatrix} & =
    \frac{2}{T} \int_{0}^{T}
    \begin{bmatrix}
    x(t)\\ y(t)
    \end{bmatrix} \sin\left({\frac{2n\pi t}{T}} \right)\mathrm{d}t. \\
    \end{split}
\end{equation} 
When $x(t)$ and $y(t)$ are segmented into $P$ points, the coefficient equations \ref{intcoeffs} are rewritten as
\begin{equation} \label{sumcoeffs}
\begin{split}
    \begin{bmatrix}
    A_{0}\\ C_{0}
    \end{bmatrix} & =
    \frac{1}{2T} \sum_{p=1}^{P}
    \begin{bmatrix}
    \Delta x_{p} \\ \Delta y_{p}
    \end{bmatrix} \frac{1}{\Delta t_{p}} \Delta (t_{p})^{2} - 
    \begin{bmatrix}
    \xi_{p} \\ \delta_{p}
    \end{bmatrix}, \\
    \begin{bmatrix}
    A_{n}\\ C_{n}
    \end{bmatrix} & =
    \frac{T}{2n^{2}\pi^{2}} \sum_{p=1}^{P}
    \begin{bmatrix}
    \Delta x_{p} \\ \Delta y_{p}
    \end{bmatrix} \frac{1}{\Delta t_{p}}
    \bigg[\cos\left({\frac{2n\pi t_{p}}{T}}\right)-\cos\left({\frac{2n\pi t_{p-1}}{T}}\right)\bigg], \\
    \begin{bmatrix}
    B_{n}\\ D_{n}
    \end{bmatrix} & =
    \frac{T}{2n^{2}\pi^{2}} \sum_{p=1}^{P}
    \begin{bmatrix}
    \Delta x_{p} \\ \Delta y_{p}
    \end{bmatrix} \frac{1}{\Delta t_{p}}
    \bigg[\sin\left({\frac{2n\pi t_{p}}{T}}\right)-\sin\left({\frac{2n\pi t_{p-1}}{T}}\right)\bigg], \\
    \end{split}
\end{equation} 
with
\begin{equation} \nonumber
\begin{split}
    \begin{bmatrix}
    \xi_{p} \\ \delta_{p}
    \end{bmatrix} =
    \sum_{j=1}^{p-1}
    \begin{bmatrix}
    \Delta x_{j} \\ \Delta y_{j}
    \end{bmatrix} - 
    \begin{bmatrix}
    \Delta x_{p} \\ \Delta y_{p}
    \end{bmatrix} \frac{1}{\Delta t_{p}} \Delta t_{j}. 
    \end{split}
\end{equation}
To calculate the coefficients such that every series is aligned to the semi-major axis, we rotate the coefficient matrix by an angle $\phi$ and a phase shift $\theta$, where the rotated coefficients are defined as 

\begin{equation} \label{rotnormcoeffs}
    \begin{bmatrix}
    A_{n}' & B_{n}' \\
    C_{n}' & D_{n}'
    \end{bmatrix} =
    \begin{bmatrix}
    A_{n} & B_{n} \\
    C_{n} & D_{n}
    \end{bmatrix}
    \begin{bmatrix}
    \cos\left({n\theta}\right) & -\sin\left({n\theta} \right)\\
    \sin\left({n\theta}\right) & \cos\left({n\theta}\right)
    \end{bmatrix}, \quad {\rm and}~~~\;
\end{equation}

\begin{equation} \nonumber
    \begin{bmatrix}
    A_{n}'' & B_{n}'' \\
    C_{n}'' & D_{n}''
    \end{bmatrix} =
    \begin{bmatrix}
    \cos\left({\phi}\right) & \sin\left({\phi} \right)\\
    -\sin\left({\phi} \right)& \cos\left({\phi}\right)
    \end{bmatrix}
    \begin{bmatrix}
    A_{n}' & B_{n}' \\
    C_{n}' & D_{n}'
    \end{bmatrix},
\end{equation}
and $\phi$ and $\theta$ are defined as 
\begin{equation} \label{efdangles}
\theta = \frac{1}{2} \tan^{-1}\left({\frac{2(A_{1}B_{1}-C_{1}D_{1})}{A_{1}^{2}+B_{1}^{2}+C_{1}^{2}+D_{1}^{2}}}\right), \qquad \phi = \tan^{-1}\left({\frac{C_{1}'}{A_{1}'}}\right).
\end{equation} \\
Furthermore, the coefficients can be size normalised by dividing them by the magnitude of the semi-major axis, 

\begin{equation} \label{normcoeffs}
    \begin{bmatrix}
    \overline{A}_{n}'' & \overline{B}_{n}'' \\
    \overline{C}_{n}'' & \overline{D}_{n}''
    \end{bmatrix} = \frac{1}{\mid {A}_{n}'' \mid}
    \begin{bmatrix}
    A_{n}'' & B_{n}'' \\
    C_{n}'' & D_{n}''
    \end{bmatrix}.
\end{equation}

\section{Hyperparameter Tuning Details}
\label{appendix: hyperparameter tuning details}

The hyperparameters tuned and their priors are given in Table \ref{tab:nn priors} for the neural network and Table \ref{tab:xgb priors} for XGBoost, with a uniform distribution applied to the priors. Hyperparameter tuning is conducted independently for each model and each combination of input features, and the optimal hyperparameter values in each of these cases are given in Table \ref{tab:xgb hp vals} for XGBoost and Table \ref{tab:nn hp vals} for the neural network. In the neural network, a preliminary run showed that batch size did not make a significant difference to the final,  so a batch size of 50 is chosen following on from \cite{E2CNN}. 
\newpage
\begin{table}
\centering
\caption{Hyperparameter priors for the neural network.}
\label{tab:nn priors}
\begin{tabular}{lll}
\hline
Hyperparameter   & Prior minimum & Prior maximum       \\ \hline
Threshold number & $1$& $254$                 \\
Order            & $1$             & $10$                  \\
Learning rate    & $\mathrm {1 \times 10^{-8}}$ & $0.01$\\
Weight decay     & $\mathrm 1 \times 10^{-10}$  & $0.001$\\
Layer size       & $10$            & $800$                 \\
Batch size       & $1$             & $50$                  \\
Dropout          & $0$            & $0.5$                 \\ \hline
\end{tabular}
\end{table}

\begin{table}
\centering
\caption{Hyperparameter priors for XGBoost.}
\label{tab:xgb priors}
\begin{tabular}{lll}
\hline
Hyperparameter       & Prior minimum & Prior maximum \\ \hline
Threshold number     & $1$             & $254$           \\
Order                & $1$             & $10$            \\
Learning rate        & $0.0001$        & $0.02$          \\
Number of estimators & $1$             & $400$           \\ \hline
\end{tabular}
\end{table}
\begin{table}
\centering
\caption{Optimal hyperparameters for every XGBoost run.}
\label{tab:xgb hp vals}
\begin{tabular}{llllllll}
\hline
Hyperparameter       & MF      & HF      & EFD     & MF, HF  & MF, EFD & HF, EFD & MF, HF, EFD \\ \hline
Threshold number     & $73$    & \textit{n/a}     & $248$   & $63$    & $192$   & $134$   & $9$         \\
Order                & \textit{n/a}     & \textit{n/a}     & $8$     & \textit{n/a}     & $1$     & $1$     & $1$         \\
Learning rate        & $0.017$ & $0.011$ & $0.019$ & $0.013$ & $0.011$ & $0.019$ & $0.015$     \\
Number of estimators & $393$   & $275$   & $333$   & $360$   & $245$   & $393$   & $295$       \\ \hline
\end{tabular}
\end{table}

\begin{table}
\centering
\caption{Optimal hyperparameters for every neural network run.}
\label{tab:nn hp vals}
\begin{tabular}{llllllll}
\hline
Hyperparameter   & MF      & HF       & EFD     & MF, HF  & MF, EFD & HF, EFD & MF, HF, EFD \\ \hline
Threshold number & $110$   & \textit{n/a} & $4$     & $2$     & $70$    & $2$     & $1$         \\
Order            & \textit{n/a} & \textit{n/a} & $3$     & \textit{n/a} & $8$     & $2$     & $51$        \\
Learning rate    & $0.00061$ & $0.0070$ & $0.00067$ & $0.0038$ & $0.00021$ & $0.0060$ & $0.00040$   \\
Weight decay     & $0.00051$ & $0.000021$ & $0.00046$ & $0.00014$ & $0.00098$ & $0.00081$ & $0.00063$   \\
Layer size       & $46$    & $364$    & $649$   & $22$    & $159$   & $547$   & $66$        \\
Dropout          & $0.026$ & $0.28$   & $0.19$  & $0.016$ & $0.047$ & $0.22$  & $0.0035$    \\
Batch size       & $50$    & $50$     & $50$    & $50$    & $50$    & $50$    & $50$        \\ \hline
\end{tabular}
\end{table}

\end{onecolumn}

\bsp	
\label{lastpage}
\end{document}